\documentclass{amsart}
\usepackage[foot]{amsaddr}
\usepackage{amsfonts}
\usepackage{amsmath}
\usepackage{amssymb}
\usepackage{mathtools}
\usepackage{color}
\usepackage{amsthm}
\usepackage{ifpdf}
\usepackage{todonotes}
\usepackage{graphicx, caption}
\usepackage{bm}
\usepackage{pdflscape,afterpage}
\usepackage[pdftex,
  	bookmarks=true,
	unicode=true,
	pdfborder={0 0 0},
	pdfstartview={FitV},
	pdfpagelayout={TwoPageRight},
	pdftitle={},
	colorlinks=false,
	linktoc=all,
	plainpages=false,
	pdfpagelabels,
	pdffitwindow=true]{hyperref}

\usepackage{todonotes}

\usepackage[left=2cm, right=2cm, top=2cm]{geometry}

\ifpdf
\else
\fi
\theoremstyle{plain}
\newtheorem{theorem}{Theorem}[section]
\newtheorem{corollary}[theorem]{Corollary}
\newtheorem{lemma}[theorem]{Lemma}

\theoremstyle{definition}
\newtheorem{assumption}[theorem]{Assumption}

\theoremstyle{remark}
\newtheorem{remark}[theorem]{Remark}
\newcommand{\R}{\mathbb{R}}
\newcommand{\N}{\mathbb{N}}

\newcommand{\f}[2]{\frac{#1}{#2}}

\renewcommand{\bar}{\overline}
\renewcommand{\tilde}{\widetilde}
\renewcommand{\hat}{\widehat}

\numberwithin{equation}{section}
\usepackage[section]{placeins}

\def\bh{\bar{h}}
\def\eps{\varepsilon}

\def\s{\sigma}

\def\h{\mathrm h}
\def\sgn{\mathrm sgn}
\def\f{\mathrm f}

\def\cg{\mathrm{g}}

\def\M{\mathcal M}

\begin{document}

\title{Short dated smile under Rough Volatility: asymptotics and numerics}

\author[Peter K. Friz]{P. K. Friz}
\address{TU and WIAS Berlin}

\author[Paul Gassiat]{P. Gassiat}
\address{U\ Paris Dauphine, PSL University}

\author[P. Pigato]{P. Pigato}
\address{U\ Rome Tor Vergata, Department of Economics and Finance 
}

\date{\today }

\begin{abstract}
In [Precise Asymptotics for Robust Stochastic Volatility Models; Ann. Appl. Probab. 2021] we introduce a new methodology to analyze large classes of (classical and rough) stochastic volatility models, with special regard to short-time and small noise formulae for option prices, using the framework [Bayer et al; A regularity structure for rough volatility; Math. Fin. 2020]. We investigate here the fine structure of this expansion in large deviations and moderate deviations regimes, together with consequences for implied volatility. We discuss computational aspects relevant for the practical application of these formulas.
We specialize such expansions to prototypical rough volatility examples and discuss numerical evidence.
\end{abstract}

\thanks{
We are grateful to C. Bayer and M. Fukasawa for discussion and to F. Bourgey and M. Pakkanen for the Python and R code for simulating the rough Bergomi model.
We thank an anonymous reviewer for several remarks that helped us to improve the paper.
PKF and PP gratefully acknowledge financial support from European Research Council Grant CoG-683164 and German science foundation (DFG) via the cluster of excellence MATH+, project AA4-2. PG acknowledges financial support from the French ANR via the project ANR-16-CE40- 0020-01} 

\maketitle

\section{Introduction}
In \cite{friz2018precise1}, precise short-time asymptotics were established for call and put option prices under stochastic volatility, under a set of abstract conditions satisfied by most classical and rough volatility (RoughVol) models. These results are refinements of large deviation statements, providing the higher order, algebraic term in an asymptotic expression, known as Laplace expansion. For RoughVol models, short dated large deviation pricing is due to Forde and Zhang \cite{forde2017asymptotics}, as is the induced implied volatility expansion (FZ expansion), which can be seen as a ``rough'' BBF (Berestycki-Busca-Florent \cite{berestycki2004computing}) formula. Our precise asymptotics provide a mechanism to compute refined implied volatility expansions, for log-strike $k_t=x t^{1/2-H}$, of the form
\begin{equation}\label{exp:ivol:intro}
\sigma_{BS}^2(t, k_t)
 =
\Sigma^2(x)+
t^{2H}a(x)
+o(t^{2H})\mbox{ as } t\downarrow 0,
\end{equation}
where the zero-order $\Sigma(x)$ term corresponds to the rough BBF formula in \cite{forde2017asymptotics}. The next-order term is seen of order $t^{2H}$ and hence increasingly important for small Hurst parameter $H$,  the basic premise of RoughVol modelling. Inclusion of this term hinges on 
an accurate evaluation of $a$. In this paper, we assume that the volatility process is of the form $\sigma(\hat{W}_t,t^{2H})$, where $\hat{W}$ is 
the Riemann-Liouville fractional Brownian motion (fBM) given by the self-similar Gaussian Volterra process in \eqref{eq:RLfbm}. It has
Hurst exponent $H\in(0,1/2]$ and it is $\rho$-correlated with the Brownian driving the asset. 

The functions $\Sigma(x)$ and $a(x)$ do not have explicit expressions and we discuss how to compute them numerically. Following \cite{forde2017asymptotics}, $\Sigma(x)$ can be computed using the Ritz method. Moreover, we propose a method for computing $a(x)$ based on a Karhunen-Loeve (KL) decomposition of the Brownian motions. (This entails a numerical approximation to an infinite-dimensional Carleman-Fredholm determinant.)

 We also derive near-the-money (meaning, as $x\to 0$) expansions of $\Sigma(x)$ and of the term structure $a(x)$ which can alternatively be used for numerics (and have the advantage of being explicit functions of model parameters). From these asymptotics, we derive consequences for at-the-money (ATM) implied skew and curvature. We also refine some moderate deviation asymptotics for call prices and implied volatilities, cf. \cite{Friz2017,bayer2017short,GULISASHVILI20203648,jacquier2020large}.

Being able to evaluate $\Sigma(x)$ and $a(x)$ allows us to test the accuracy of the short-time asymptotics in practice. We do so with a numerical case study of the rough Bergomi (rBergomi) model. To exploit our general framework we look at a volatility given by 
\begin{equation}\label{eq:vol:general:rbergomi}
\sigma(\hat{W}_t,t^{2H})=\sigma_0\exp\left(\frac{\eta}{2}\hat{W}_t-\frac{\theta \eta^2}{4}t^{2H}\right),
\end{equation}
so that for $\theta=1$ we get the rBergomi model considered in \cite{bayer2016pricing,BLP15} with constant forward variance, for $\theta=0$ the rBergomi version in \cite{bayer2017short,forde2017asymptotics}. Note, however, that \eqref{eq:vol:general:rbergomi} is a genuine rBergomi model for any value of $\theta$, as discussed in Remark \ref{rem:fw}.
We compare our approximation to the FZ expansion from \cite{forde2017asymptotics} and to the Edgeworth asymptotics in \cite{euch2018short}. 
We consider how smiles vary as $\theta$ varies in \eqref{eq:vol:general:rbergomi} and as expiry $t$ increases. We discuss and test the volatility term structure and its slope ATM, and observe how the term $a(x) t^{2H}$ improves the asymptotics as $H$ decreases.  
We observe the same feature when we implement the moderate deviation asymptotics for implied volatility, where for $H$ small the inclusion of the term structure correction $a(0) t^{2H}$ significantly improves on the numerical results presented in \cite{bayer2017short}.

Proofs rely on stochastic Taylor expansions, rate function representations in \cite{forde2017asymptotics,bayer2017short} and on the local analysis on the Wiener space introduced in \cite{friz2018precise1,bayer2017regularity}. The classical Gao-Lee results \cite{gaolee} are used to go from option prices to implied volatility asymptotics both in large and moderate deviation regimes.

\medskip

\noindent{\bf Rough Volatility.} 
It has been shown in recent years that RoughVol models
provide great fits to observed volatility surfaces \cite{bayer2016pricing} capturing fundamental stylized facts of implied volatility in a parsimonious way. Specifically, this class of models can reproduce the steep short end of the smile, displaying exploding implied skew  \cite{alos2007short,fukasawa2011asymptotic,fukasawa2017short}, and they are the only models consistent with the power law of the skew \cite{bayer2016pricing,lee} not admitting arbitrage \cite{fukasawa2020}. RoughVol is also supported by statistical and time series analysis \cite{gatheral2018volatility,FTW19,BLP16} and by market microstructure considerations \cite{el2018microstructural}. Many authors have even argued that $H\approx 0$, such as to be consistent with a skew explosion close to $t^{-1/2}$ \cite{bayer2016pricing,BHP20}. One main aspect  of RoughVol is non-Markovianity. This is a serious complication when it comes to pricing, as Monte Carlo methods become more expensive and PDE methods are not available. For this reason, efficient simulation schemes have been proposed \cite{bayer2020,BLP15,McCrickerdPakkanen2018}. Fourier based methods are available for the rough Heston model \cite{euch2016characteristic}. Deep and machine learning approaches have also recently been discussed in \cite{bayer2019deep,goudenege}. 
Small maturity approximations are used in this context to obtain starting points for calibration procedures, which are then based on numerical evaluations.

\medskip

\noindent{\bf Asymptotic option pricing.} Classical motivation for (semi-closed form) asymptotic pricing includes fast calibration, and a quantitative understanding of the impact of 
model parameters on 
 relevant quantities such as implied skew and curvature/convexity along the moneyness dimension or slope along the term-structure dimension.
 Explicit expressions for such quantities (that follow in this setting from our expansion)
 and their shape characteristics are also used to choose the most appropriate model to be fitted to data \cite{aitsahalia2019}, leave alone being the origin of some widely used parametrisations of the volatility surface. An interesting, if recent, addition to this list comes from a machine learning perspective: the form of an expansion such as (\ref{exp:ivol:intro}) may be viewed as {\em expert knowledge}, 
which significantly narrows the learning task to finer information such as the error in that expansions; it is equally conceivable to learn $a=a(x)$ and other components in the expansion.

 Under Markovian stochastic volatility, expansion \eqref{eq:vol:general:rbergomi} is analogous, e.g., to the result derived in \cite{forde2012small} for the Heston model. There, the term structure is  $a(x)t$ (due to the diffusive scaling of the volatility), whereas here the correction term is $a(x)t^{2H}$ (due to the rough scaling of the volatility). 
Similar expansions are derived also in \cite{osajima2015general}, for more general Markovian models, and (formally) in \cite{MS03,MS07} for Markov stochastic volatility models with jumps.

In recent years several authors have studied the short-time behavior of RoughVol models. Theoretical results on short-time skew and curvature are given in \cite{fukasawa2017short,alos2017curvature}.
A second order short-time expansion is given in \cite{euch2018short} for general (rough) stochastic volatility models. In \cite{jacquier2017pathwise}, the pathwise large deviation behavior under rBergomi dynamics is studied.
Pathwise large and moderate deviation principles for (possibly rough) Gaussian stochastic volatility models are established in  \cite{GULISASHVILI20203648,gulisashvili2020timeinhomogeneous}, together with asymptotic results at the central limit (Edgeworth) regime. For the rough Heston model, the recent work \cite{Gerhold2019} provides call expansions of the same type as ours, involving the energy function and the first order algebraic term, at the same large deviations regime $k_t=x t^{1/2-H}$. (The rigid infinite-dimensional affine structure which underlies \cite{Gerhold2019} is not available for rBergomi type models as considered in this work.) As already mentioned, our work builds on the  large deviations principle proved  in \cite{forde2017asymptotics} for models with volatility $\sigma(\hat{W}_t)$, and on \cite{bayer2017short}, where the at-the-money behavior of the Forde-Zhang rate function is used to prove moderate deviation priciples and implied volatility asymptotics for the same type of models. The theoretical foundations of the present paper are given in \cite{friz2018precise1}.

In Section \ref{sec:preliminary} we explain our RoughVol setting. In Section \ref{sec:results}  we state and comment our results. In Section \ref{sec:bergomi} we discuss and implement our results in the case of the rBergomi model. In Section \ref{sec:projections} we show how $\Sigma$ and $a$ can be computed using Ritz method and KL decomposition. We collect all the proofs in Section \ref{sec:proofs}.

\section{Preliminaries on rough volatility}\label{sec:preliminary}

We consider the following RoughVol model, with $H \in (0,1/2]$, normalized to rate $r=0$ and $S_0=1$
\begin{equation}
\label{eq:S:roughvolmodel}
\frac{d S_t}{S_t} = \sigma(\hat{W}_t,t^{2H}) d (\rho W_t +  \bar{\rho} \bar W_t),
\end{equation}%
where $W, \bar{W}$ are independent Brownian motions (BM) and $\rho\in(-1,1)$, $\rho^2+\bar{\rho}^2=1$. We also write $\tilde{W} = \rho W + \bar{\rho} \bar{W}$.  Moreover, $\hat{W}=(\hat{W}_t)_{t\geq 0}$ is a Gaussian Volterra process of the form 
\begin{equation}
\label{def:fBM}
\hat W_t = (K * \dot W)_t = \int_0^t K(t,s) dW_s,
\end{equation}
for a kernel $K(t,s)$ such that
$\hat W$ is self-similar with exponent $H \in (0,1/2]$, meaning
\begin{equation}\label{eq:scaling:fBM}
\mathrm{Law}(\hat W_{\eps^2 t}: t \le \bar t) = \eps^{2 H} \mathrm{Law}(\hat W_t: t \le \bar t),
\qquad \mbox{ for some } \bar t > 0.
\end{equation}
The BM  $W$ drives the stochastic ``rough'' volatility, meaning (with abusive notation) that $\sigma(t,\omega) = \sigma(\hat{W}_t,t^{2H})$, where $\sigma(x,y)$ is a smooth deterministic real-valued function. 
We denote $\sigma'(x,y)=\partial_x \sigma(x, y)$,
$\sigma''(x,y)=\partial_{xx} \sigma(x, y)$,
$\dot{\sigma}(x,y)=\partial_y \sigma(x, y)$. We also denote $\sigma_0=\sigma(0,0)$ $>0$ the spot volatility and
\begin{equation}
\sigma_0'=\sigma'(0, 0),\quad\sigma_0''=\sigma''(0,0),\quad\dot{\sigma}_0=\dot \sigma(0, 0),
\end{equation} 
the derivatives of the volatility function at the initial condition. 
We consider a dependence in $t^{2H}$ in $\sigma(\cdot)$, because this is the scaling of the variance of the fBm at time $t$.  For this reason, this is the scaling of the time-dependent term in the  rBergomi model, and also the scaling such that we observe a dependence in $\dot{\sigma}_0$ in our precise asymptotics.  We apply the abstract results proved in \cite{friz2018precise1} for $K(t,s) = const \times (t-s)^{H-1/2}$. However, we expect these approximations to hold in greater generality:
the same type of expansions
should hold for other kernels such that $\hat W$ in \eqref{def:fBM} satisfies \eqref{eq:scaling:fBM}.
Self-similarity is equivalent to the fact that $K$ can be written in the following form
\begin{equation}\label{eq:scaling:K}
K(t,s) = (t-s)^{H-1/2} f_K(s/t),
\end{equation}
for a suitable function $f_K$ (see \cite[Lemma 2.4]{jost2007}), so that all such kernels can be seen as a perturbation of $(t-s)^{H-1/2}$.  Two  classical processes of this form are the Mandelbrot-Van Ness and the Riemann-Liouville fBMs (see Appendix \ref{sec:fBM}). Without loss of generality, we also assume $K(t,s)=0$ for $t<s$. 

A similar setting has been considered in \cite{forde2017asymptotics,bayer2017short}. The main difference in the structure of the model is that here we allow for a direct dependence on time in $\sigma(t,\omega) = \sigma(\hat{W}_t,t^{2H})$, whereas in \cite{forde2017asymptotics,bayer2017short}  the volatility function depends only on the fBM, so  $\sigma(t,\omega)=\sigma(\hat{W}_t)$. As mentioned in the introduction, assuming that the volatility is a deterministic function only of the fBM rules out the rBergomi model $\sigma(\hat{W}_t,t^{2H})=\sigma_0 \exp(\eta\hat{W}_t/2-\eta^2 t^{2H}/4)$, see \cite{bayer2016pricing,BLP15}, from the analysis, so a modified version of rBergomi is  considered in \cite{bayer2017short}. We discuss in detail both versions of this model in Section  \ref{sec:bergomi}.
With a volatility function $ \sigma(\hat{W}_t,t^{2H})$, one can write the dynamics of the log-price $X=\log S$ as
\begin{equation}\label{eq:roughvolmodel}
X_t=\int_{0}^t \s \left( \hat W _s,s^{2H}\right) 
d ( \bar{\rho} \bar W+\rho W)_s -\frac{1}{2}\int_{0}^t
\s^2 (  \hat W_s,s^{2H}) ds.
\end{equation}%
In this case, a LDP holds, writing $\hat\varepsilon=\varepsilon^{2H}$, for  
\begin{equation}\label{eq:rescaled:log:price}
\bar X^\eps_1 = \int_0^1 \s \left( \hat \eps \hat W_t, \hat\varepsilon^2 t^{2H} \right) \hat\varepsilon
d (\bar{\rho} \bar W+\rho W)_t -\frac{1}{2}\varepsilon \hat{\eps} \int_0^1
\s^2 \left(  \hat \eps \hat W_t,\hat\varepsilon^2 t^{2H}\right) dt \ ,
\end{equation}
with speed $\hat{\varepsilon}^{2}$ and rate function
\begin{equation}\label{ifzt}
          \Lambda(x)  := \inf \left\{ \tfrac{1}{2}\| h,\bh \|^2_{H^1} :  \int_{0}^1 \s \big(\, \hat h , 0 \, \big) 
d\left( \bar{\rho} \bh+\rho h \right) = x \right\} \equiv \tfrac{1}{2}\| h^x,\bh^x \|^2_{H^1} \ ,
\end{equation}      
where $\hat h_t = (K * \dot h)_t $ and $\| \cdot \|_{H^1}$ is the Cameron-Martin norm. The existence of a minimizer above is obtained from a standard compactness argument. Through the space-time scaling $t=\eps^2$ and the fact that, in law, $\bar X^\eps_1=\frac{\hat{\eps}}{\eps} X_{\eps^2}$, this small-noise LDP translates to a short-time LDP.
This result was proved for $ \sigma(\hat{W}_t,t^{2H})= \sigma(\hat{W}_t)$ in \cite{forde2017asymptotics} and then extended to possible dependence in $t^{2H}$ in \cite[Section 7.3]{friz2018precise1}.
In general, when looking only at large (or moderate) deviations, the $t^{2H}$-dependence in $\sigma(\cdot)$ does not affect the analysis, and the large (or moderate) deviations behavior is the same one would get with volatility $ \sigma(\hat{W}_t,0)$.
In \cite{friz2018precise1}, we consider a general asymptotic setting, obtaining for generic stochastic volatility models (including RoughVol ones) precise asymptotics that refine such large deviations asymptotics. For such refinement, this $t^{2H}$-dependence actually affects the asymptotics. In the present paper we provide computationally relevant results that allow for the practical usage of such refined pricing asymptotics and discuss their consequences  on the Black-Scholes implied volatility.

\section{Results} \label{sec:results}

We consider call and put prices under model \eqref{eq:roughvolmodel}, i.e.
\[
\begin{split}
c(t,k) = E[(\exp X_t-\exp k )^+],\quad\quad\quad\quad
p(t,k) = E[(\exp k - \exp X_t)^+],
\end{split}
\]
where $k$ is the log-strike (or  log-moneyness).
In \cite[Theorem 1.1]{friz2018precise1} we obtain precise small-noise price expansions for generic (classical and rough) volatility dynamics.
As in the classical Brownian case, such small noise results can be translated into short-time results writing $t=\eps^2$. In this paper, we focus on the short-time setting. We write $\sim$ for asymptotic equivalence, $f_t\sim g_t$ if $f_t/g_t\to 1$ as $t\to 0$, and ``$\approx$'' for ``is close to'' in informal terms. 
We also write $\sigma_x^2=2\Lambda(x)/\Lambda'(x)^2$. 
\begin{assumption}
Throughout the paper,  we assume $K$ in \eqref{def:fBM} is of the form
\[
K(t,s) = const \times (t-s)^{H-1/2}.
\]
\end{assumption}
In short-time, \cite[Theorem 1.1]{friz2018precise1} reads as follows:

\begin{theorem} \label{thm:main:smalltime} Let $H \in (0,1/2]$ and $k_t = x  t^{1/2-H} $. 
Assume that a LDP holds for $c,p$ above, 
and the existence of $1^+$ moments for $\exp X_t$. Then, for $x>0$ small enough, the rate function $\Lambda = \Lambda(x)$ is continuously differentiable at $x$ and
\begin{equation*}\label{eq:lim:callprice:short:time}
       c(t,k_t) \sim \exp \left( {  - \frac{\Lambda(x)}{t^{2H}}  } \right)  t^{1/2+2H} \frac{A(x)}{ (\Lambda'(x))^2  \sigma_x \sqrt{2\pi}} \ \ \text{as $t \downarrow 0$,}
\end{equation*} 
for some function $A(x)$ with $A(x) \to 1$ as $x \downarrow 0$. Similarly, for
$x  < 0$, close enough to $0$, we have
\begin{equation*}\label{eq:lim:putprice:short:time}
       p(t,k_t)\sim \exp \left( {  - \frac{\Lambda(x)}{t^{2H}}  } \right)  t^{1/2+2H} \frac{A(x)}{ (\Lambda'(x))^2  \sigma_x \sqrt{2\pi}} \ \ \text{as $t \downarrow 0$,}
\end{equation*} 
for some function $A(x)$ with $A(x) \to 1$ as $x \uparrow 0$. Moreover, such $A$ can be expressed as
\begin{equation} \label{Aabstract}
 A(x) = 
\begin{cases}
    E \big[ \exp ( \Lambda'(x) \Delta_2^x ) \big],& \text{if } H < 1/2,\\
    e^x E \big[ \exp ( \Lambda'(x) \Delta_2^x ) \big],& \text{if } H=1/2,
\end{cases}
\end{equation}
where $\Delta_2^x$ is a certain quadratic Wiener functional (specified in \cite[Equation (7.4)]{friz2018precise1}, see also \eqref{def:Delta2} below).
\end{theorem}
\begin{remark}The fact that $x>0$ above has to be taken small enough is in order for the minimizer $(h^x,\bh^x)$ in \eqref{ifzt} to be unique and non-degenerate. The latter means, in a nutshell, that the Hessian of $I(h, \bh) := \tfrac{1}{2}\| h,\bh \|^2_{H^1}$ is strictly positive when restricted to those $(h,\bh)$ such that $\int_{0}^1 \s \big(\, \hat h , 0 \, \big) 
d\left( \bar{\rho} \bh+\rho h \right) = x$, and is equivalent to the finiteness of $A(x)$ defined above.
\end{remark}
We write $Kf(t)=\int_0^t K(t,s)f(s)ds$, 
$K^2f(t)=\int_0^t K^2(t,s)f(s)ds$
and $\langle \cdot, \cdot\rangle$ for  the inner product in $L^2[0,1]$. We also denote $\bar{K}$ the adjoint of $K$ in $L^2[0,1]$ so that $\bar{K}1(u)=\int_u^1 K(t,u) dt$. 
Fully explicit expressions are computable in the case of the Riemann-Liouville fBM (Appendix \ref{sec:fBM}) and in particular in the case of standard BM (this is the classical case of Markovian stochastic volatility). We denote
\[
\begin{split}
C_{K,\rho}&=\frac{\langle K^2 1, 1\rangle}{2}-\frac{3}{2} \langle (\bar{K}1)^2,1 \rangle
+
 \rho^2\bigg(
\frac{17}{2} \langle K1,1\rangle^2
- \frac{3}{2}\langle (K1)^2,1 \rangle
-  3 \langle K1,\bar{K}1 \rangle
 \bigg)
,
\\
\bar{C}_{K,\rho}&=\frac{\langle K^2 1, 1\rangle}{2}
-
\frac{3}{2}\rho^2  \langle (K1)^2,1 \rangle.
\end{split}
\]

\begin{lemma}[Fine structure of $A$]\label{expansion:A}
 For $H\in (0,1/2]$, the following expansion holds for $A(x)$ as $x\rightarrow 0$:
\begin{equation}\label{expansion:A:rough}
A(x)= 
1- x \frac{\rho \sigma_0' \langle K1,1\rangle}{ \sigma_0^2}+x^2 
\bigg( \frac{(\sigma_0')^2}{\sigma_0^4}  C_{K,\rho}+
 \frac{\sigma''_0}{\sigma_0^3}
\bar{C}_{K,\rho}+  \frac{\dot\sigma_0}{(2H+1) \sigma_0^3}  \bigg)
+
\left(\frac{x}{2}+
\frac{x^2}{8}\right) {\bf 1}_{\{H=1/2\}}
 +O(x^3).
\end{equation}
\end{lemma}
As a consequence of Theorem \ref{thm:main:smalltime} the following expansion holds for the Black-Scholes implied volatility (by a standard application of Gao-Lee \cite{gaolee}, detailed in \cite[Appendix D]{friz2018precise1}).
\begin{corollary}[Asymptotic smile and term structure at the large deviations regime]
\label{corollary:ivol} 
Writing $k_t=xt^{1/2-H}$, we have the following expansion, for $x\in \R \setminus \{0\}$ such that Theorem \ref{thm:main:smalltime} holds:
\begin{equation}\label{ivol:expansion}
\sigma_{BS}^2(t, k_t)
 =
\Sigma^2(x)+
t^{2H}a(x)
+o(t^{2H})\mbox{ as } t\downarrow 0,
\end{equation}
where
\begin{equation}\label{def:v}
\Sigma(x)=\frac{|x|}{\sqrt{2\Lambda(x)}}
\end{equation}
and
\begin{equation}\label{def:a}
a (x)
=
\begin{cases}
\frac{
x^2 }{
2\Lambda(x)^2
  }
 \log \bigg( \frac{ 2A(x)\Lambda(x) }{ \Lambda'(x) x }
\bigg) & \mbox { if } H<1/2, \\
\frac{
x^2 }{
2\Lambda(x)^2
  }
 \log \bigg( \frac{ 2A(x)\Lambda(x) }{ \Lambda'(x) x \exp(x/{2})}
\bigg) & \mbox { if } H = 1/2.
\end{cases}
\end{equation}

\end{corollary}
\begin{remark}
In general, from a LDP for call prices follows the celebrated BBF formula for implied volatility (Berestycki-Busca-Florent \cite{berestycki2004computing}, see also  also Pham \cite{pham2010large} for a derivation). Under RoughVol pricing with $\sigma(\omega,t)=\sigma(\hat W_t)$, this has been extended in \cite{forde2017asymptotics} to
\begin{equation}\label{bbf}
\sigma_{BS}^2(t,k_t)\sim \frac{x^2}{2\Lambda(x)},
\end{equation}
holding  for fixed $x$, in short-time, with $k_t=xt^{1/2-H}$.
Thanks to the $A$-term in \eqref{eq:lim:callprice:short:time}, we can extend this approximation, adding the term structure $t^{2H}a(x)$.
Note that the expansions hold for $H\in (0,1/2]$, but for $H=1/2$ their functional form is different, as some additional terms appear in $A(x)$ and in the term structure of the Black-Scholes implied volatility $a(x)$. 
\end{remark}

We denote now
\begin{equation}\label{def:D}
\begin{split}
D_{K,\rho}&= 
\langle K^2 1, 1\rangle-\langle (\bar{K}1)^2,1 \rangle
+
 \rho^2\big(
3 \langle K1,1\rangle^2
- \langle (K1)^2,1 \rangle
-  2 \langle K1,\bar{K}1 \rangle
 \big),
 \\
\bar{D}_{K,\rho}&= \langle K^2 1, 1\rangle
-
\rho^2  \langle (K1)^2,1 \rangle.
\end{split}
\end{equation}
The short-time implied volatility coefficients in the previous statement can be expanded as follows near-the-money.
\begin{theorem}[At-the-money expansion of the coefficients]\label{expansion:atm:ivol}
For
$x\rightarrow 0$, the $\Sigma$ coefficient has the following expansion:
\begin{equation}\label{expansion-v}
\begin{split}
\Sigma(x)&= \sigma_0+
x \Sigma'(0)
+ x^2 \frac{\Sigma''(0)}{2} +O(x^3),
\end{split}
\end{equation}
where
\[
\begin{split}
\Sigma'(0)&=\frac{\rho \sigma_0'\langle K1,1\rangle}{\sigma_0}, \\
\frac{\Sigma''(0)}{2}
&=
 \frac{(\sigma_0')^2}{\sigma_0^3}
\left\{
- 3 \rho^2 
 \langle K1,1\rangle^2
+ \frac{\rho^2}{2}
 \langle (K1)^2 ,1\rangle  
+
\frac{ 1}{2}
 \langle (\bar{K}1)^2 ,1\rangle  
+
 \rho^2 
\langle K1,\bar{K}1\rangle
\right\}
+ \frac{\sigma_0''}{\sigma_0^2}  \frac{ \rho^2 }{2}
 \langle (K1)^2 ,1\rangle.  
\end{split}
\]
The term structure coefficient, at the first order in $x$ at $0$, is
\begin{equation}\label{expansion-a}
\begin{split}
a(x)  =a_0+O(x),
\end{split}
\end{equation}
with
\[
a_0=
(\sigma_0')^2 D_{K,\rho}+
  \sigma_0 \sigma''_0
 \bar{D}_{K,\rho}+
 \frac{ \sigma_0 \dot\sigma_0}{H+1/2}
 +\rho \sigma_0' \sigma_0^2 \langle K1,1\rangle {\bf 1}_{\{H=1/2\}} \,.
\]
\end{theorem}
\begin{remark}
From definition \eqref{def:v}-\eqref{def:a} and from the fact that $\Lambda$ is quadratic in $x$ we see that \eqref{expansion-a} implies a  relation between $A$ and $\Lambda$ for $x\rightarrow 0$.
\end{remark}
\begin{remark}
Implied variance expansion \eqref{ivol:expansion} reads as follows on implied volatility
\begin{equation}\label{exp:implied_vol_KL}
\sigma_{BS}(t,k_t)\approx
\frac{|x|}{\sqrt{2\Lambda(x)}}
+t^{2 H} 
\frac{a(x)}{|x|}\sqrt{\frac{\Lambda(x)}{2}}.
\end{equation}
In order to implement these expansions, one can use the methods discussed in Section \ref{sec:projections}, computing numerically the rate function $\Lambda(x)$ and $\Sigma(x)$ using FZ expansion, and then computing $a(x)$ using KL. 
However, this last step can be computationally expensive, since a large number of basis functions are needed for the KL decomposition to be accurate, for $H$ close to $0$. As an alternative, one can use approximation 
\begin{equation}\label{exp:implied_variance_rate_function}
\sigma_{BS}(t,k_t)\approx
 \Sigma(x) +t^{2 H} \frac{a_0}{2\sigma_0} 
=
\frac{|x|}{\sqrt{2\Lambda(x)}}
+t^{2 H} \frac{a_0}{2\sigma_0},
\end{equation}
 for implied volatility, which follows from implied variance expansion \eqref{ivol:expansion} and \eqref{expansion-a}.
If the rate function cannot be computed, we can use \eqref{expansion-v} to expand the implied volatility as
\begin{equation}\label{exp:implied_variance}
\sigma_{BS}(t,k_t)\approx 
\Sigma(0)+ \Sigma'(0)x+ \frac{\Sigma''(0)}{2}x^2
+t^{2 H} \frac{a_0}{2\sigma_0}
 .
\end{equation}
In particular, we get the following explicit expansion for the ATM term structure:
\begin{equation}\label{expansion:ATM:ivol}
\sigma_{BS}(t, 0)
=
\sigma_0+
t^{2H}\frac{a_0}{2\sigma_0} 
+o(t^{2H}).
\end{equation}
\end{remark}

\begin{remark}[The term structure of implied volatility]\label{rem:empirical:term structure}
From the expansion of the ATM term structure \eqref{expansion:ATM:ivol} we also see, in the short end, that ${\sigma}_{BS}^2(t,0)$ is increasing in $t$ if $a_0>0$ and decreasing if $a_0<0$. 
This may be compared with a large body of literature concerning monotonicity properties of the term structure of implied volatility, see e.g. \cite{
CAMARA2011215,GuoHanZhao14,KrylovaNikkinenVahamaa09,Vasquez}.
\end{remark} 

\begin{corollary}[Skew and curvature at the large deviation regime]
Let $k_t=x t^{1/2-H}$, for $x\in \R  \setminus \{0\}$. Then, if $H<1/2$, for $t\downarrow 0$

\begin{eqnarray}
\frac{
\sigma_{BS}(t, k_t)
-
\sigma_{BS}(t, -k_t)
}
{2 k_t}
&\sim & \frac{\Sigma(x)-\Sigma(-x)}{2x} t^{H-1/2} \,.\\
\frac{
\sigma_{BS}(t, k_t)
+
\sigma_{BS}(t, -k_t)
-2
\sigma_{BS}(t, 0)}
{k_t^2}
&\sim & 
\frac{
\Sigma(x)+
\Sigma(-x)-2
\Sigma(0)}{x^2}
t^{2H-1}
\,.\end{eqnarray}
\end{corollary}

\begin{remark}
The quantities in the rhs of the equivalences converge as $x\downarrow 0$ to  $\Sigma'(0),\Sigma''(0)$ given in Theorem \ref{expansion:atm:ivol}.
The quantities in the lhs of the equivalences are finite difference approximations of ATM implied volatility skew $\partial_k \sigma_{BS}(t,0)$ and curvature $\partial_{kk}\sigma_{BS}(t,0)$ .
Such finite differences are relevant because only a finite number of prices are observable on real markets. 
They give skew and curvature at the large deviation regime, a result that complements \cite{fukasawa2017short,euch2018short}  (skew and curvature at central limit  regime), \cite{bayer2017short} (skew at moderate deviation regime),
 \cite{Gerhold2019} (skew and curvature at large deviations regime for rough Heston), \cite{alos2017curvature} (true skew and curvature).

From these formulas, we also infer  the sign of implied skew and of implied curvature (convexity). Indeed, if $\sigma_0,\sigma_0'\neq 0$, it is clear that  $\sgn(\Sigma'(0))=\sgn(\rho)$ and that
\begin{equation}\label{eq:sign:curvature:general}
\begin{cases}
\Sigma''(0)=0\quad
\mbox{ iff }
\quad
 \rho^2 
 =
\frac{   \langle
(\bar{K}1)^2 ,1\rangle  }{
6
 \langle K1,1\rangle^2
- 
 \langle (K1)^2 ,1\rangle  
- 2
\langle K1,\bar{K}1\rangle
 - \frac{\sigma_0''\sigma_0}{(\sigma_0')^2}  
 \langle (K1)^2 ,1\rangle
},
\\
\Sigma''(0)<0\quad
\mbox{ iff }
\quad
\rho^2 
 >
\frac{   \langle
(\bar{K}1)^2 ,1\rangle  }{
6
 \langle K1,1\rangle^2
- 
 \langle (K1)^2 ,1\rangle  
- 2
\langle K1,\bar{K}1\rangle
 - \frac{\sigma_0''\sigma_0}{(\sigma_0')^2}  
 \langle (K1)^2 ,1\rangle
} > 0,
\\
\Sigma''(0)>0\quad
\mbox{ otherwise.}
\end{cases}
\end{equation}
\end{remark}

\begin{theorem}[Moderate deviations]\label{th:moderate_deviations}
Assume that $\Lambda$ is $i\in \N$ times continuously differentiable.
Let $H\in (0,1/2)$, $\beta>0$ and $n\in \N$ such that $ \beta \in(\frac{2H}{n+1}, \frac{2H}{n}]$.
Set $k_t=x t^{1/2-H+\beta}$. Then
\[
c(t,k_t)\sim \exp\left(-
\sum_{i=2}^{n} \frac{\Lambda^{(i)}(0)}{i!} x^i t^{i\beta-2H}
\right)t^{1/2+2H-2\beta}\frac{\sigma_0^3}{x^2 \sqrt{2\pi}}.
\]
Moreover
\begin{equation}\label{exp:imp:vol}
\begin{split}
\sigma_{BS}^2(t, k_t)=
\sum_{j=0}^{n-2}
(-1)^j 2^j \sigma_0^{2(j+1)}
 \left(
\sum_{i=3}^{n} \frac{\Lambda^{(i)}(0)}{ i!} x^{i-2} t^{(i-2)\beta}
 \right)^j
+o(t^{2H-2\beta}).
\end{split}
\end{equation}

\end{theorem}

\begin{remark}\label{rem:moderate_deviations}
An implied volatility expansion similar to \eqref{exp:imp:vol} was proved in \cite{bayer2017short}, in the case $\sigma(t,\omega)=\sigma(\hat{W}_t)$, for $ \beta \in [\frac{2H}{n+1}, \frac{2H}{n})$,
with remainder of order $\max(t^{2H-2\beta-\eps},t^{(n-1)\beta})$. The derivatives of the rate function were computed until $\Lambda'''(0)$, here we also computed $\Lambda^{(4)}(0)$ (cf. Lemma \ref{expansion:J}). This allows us to use the second order moderate deviation (instead of first order as in \cite{bayer2017short})
\[
\sigma(t,k_t)=\Sigma(0)+\Sigma'(0)x t^\beta + \frac{\Sigma''(0)}{2}x^2 t^{2\beta}+o(t^{2H-2\beta})
\,.
\]
Moreover, even if it does not show up in the asymptotics, the term structure can be incorporated as follows
\[
\sigma(t,k_t)\approx \Sigma(0)+\Sigma'(0)x t^\beta + \frac{\Sigma''(0)}{2}x^2 t^{2\beta}+\frac{a_0}{2\sigma_0}t^{2H},
\]
and this provides a sensible improvement in the implementation of such short-time result (cf. Figure \ref{fig:mod:dev}).
\end{remark}

\section{A case study: the rough Bergomi model}\label{sec:bergomi}

\subsection{The rough Bergomi model}

 Introduced in \cite{bayer2016pricing}, as a modification of the classical Bergomi model where the exponential (Ornstein-Uhlenbeck) kernel is replaced by a power-law kernel, the rBergomi model provides great fits of empirical implied volatility surfaces with a very small number of parametres. In such model, the volatility is given by the ``Wick'' exponential of a Riemann-Liouville fBM
\begin{equation}\label{eq:vol:genuine:rbergomi}
\sigma(t,\omega)=\sigma_0\exp\left(\frac{\eta}{2}\hat{W}_t-\frac{\eta^2}{4}t^{2H}\right).
\end{equation}
In the most general framework \cite{bayer2016pricing}, the constant $\sigma^2_0$ is replaced by the forward variance curve, which is a function of time observable on the market (so it plays the role of an initial condition, cf. also Remark \ref{rem:fw}). 
The specific volatility in \eqref{eq:vol:genuine:rbergomi} did not fit in the framework of \cite{forde2017asymptotics, bayer2017short}, as in these papers the volatility is assumed to be $\sigma(\hat{W}_t)$. For this reason, in \cite{bayer2017short}, the following version of the rBergomi  model is considered
\begin{equation}\label{eq:vol:modified:rbergomi}
\sigma(t,\omega)=\sigma_0\exp\left(\frac{\eta}{2}\hat{W}_t\right).
\end{equation}
In this work we consider \eqref{eq:vol:general:rbergomi}, a version of the rBergomi model with one additional parameter $\theta \in \R$, that includes both the previous ones (for $\theta=0,1$). The volatility function in \eqref{eq:S:roughvolmodel} is 
\begin{equation}\label{eq:funct:general:rbergomi}
\sigma(x,y)=\sigma_0\exp\left(\frac{\eta}{2}x-\frac{\theta \eta^2}{4}y\right)
\,.
\end{equation} 
The interpretation of the parameters is the following: $\sigma_0$ is the spot volatility and $\eta$ represents the volatility of volatility. The parameters of the driving noise are the Hurst exponent $H$ of $\hat{W}$ and the correlation parameter $\rho$ between the BM $\tilde{W}$ driving the asset and $W$ in \eqref{def:fBM}. We can interpret the newly introduced $\theta$ parameter as a damping coefficient of the volatility.
\begin{remark}\label{rem:fw}
Note that the forward variance curve model 
$\xi_t(u)=E[ \sigma^2(\hat{W}_u,u^{2H}) |\mathcal{F}_t]$  (and $\mathcal{F}_t$
filtration generated by $W$)
induced by \eqref{eq:funct:general:rbergomi} and \eqref{eq:vol:general:rbergomi},
 is a genuine rBergomi model for any value of $\theta$, with different values of $\theta$ corresponding to different specifications of the initial variance curve. More precisely, for fixed $\theta$,
\[
\xi_0(u)=
E[ \sigma^2(\hat{W}_u,u^{2H})]
=
\sigma_0^2 \exp\left(\frac{(1-\theta) \eta^2}{2}u^{2H}\right).
\]
\end{remark}
Coming now to short-time pricing, Lemma \ref{expansion:J} holds for the general model in \eqref{eq:vol:general:rbergomi}, so that we are able to compare our asymptotics with  large or moderate deviations results for the different versions of rBergomi in \cite{bayer2017short,forde2017asymptotics,jacquier2017pathwise}.
However, in Corollary \ref{corollary:ivol}, $\Sigma^2(x)$ is not affected by the value of $\theta$, but the term structure $a(x)$ is. 

From the volatility function \eqref{eq:funct:general:rbergomi} we get
\[
\begin{split}
\sigma_0=\sigma_0, \quad
\sigma_0'= \frac{ \sigma_0 \eta}{2}, \quad
\sigma_0''=  \frac{ \sigma_0 \eta^2}{4},
\quad \dot\sigma_0=- \frac{\theta \eta^2}{4}\sigma_0,
\end{split}
\]
so all constants can be simplified. In particular condition \eqref{eq:sign:curvature:general} for the convexity of the short-time smile (with $\sigma_0,\eta\neq 0)$ simplifies to
\[
\begin{cases}
\Sigma''(0)=0\quad
\mbox{ iff }
\quad
 \rho^2 
 =
\frac{   \langle
(\bar{K}1)^2 ,1\rangle  }{
6
 \langle K1,1\rangle^2
- 
2 \langle (K1)^2 ,1\rangle  
- 2
\langle K1,\bar{K}1\rangle
},
\\
\Sigma''(0)<0\quad
\mbox{ iff }
\quad
\rho^2 
 >
\frac{   \langle
(\bar{K}1)^2 ,1\rangle  }{
6
 \langle K1,1\rangle^2
- 2 
 \langle (K1)^2 ,1\rangle  
- 2
\langle K1,\bar{K}1\rangle
} > 0,
\\
\Sigma''(0)>0\quad
\mbox{ otherwise.}
\end{cases}
\]
(note the dependence only on $H$, through $K$, and $\rho$).  
On calibrated parameters (for example in \cite{bayer2016pricing}) we have that the condition for vanishing second derivative is almost satisfied. 
This means that the short-time ATM curvature is very close to $0$, and indeed observed smiles are almost linear ATM. 

All the constants in previous expansions depend on the kernel $K$. For the Riemann-Liouville kernel \eqref{kernel:RLfBM} the $K$-functionals involved are explicit, given in \eqref{explicit:fractional}.

\subsection{Implementation of Rough Bergomi}\label{sec:implementation}

\begin{figure}[ht!]
    \begin{center}
	\includegraphics[width=165mm]	{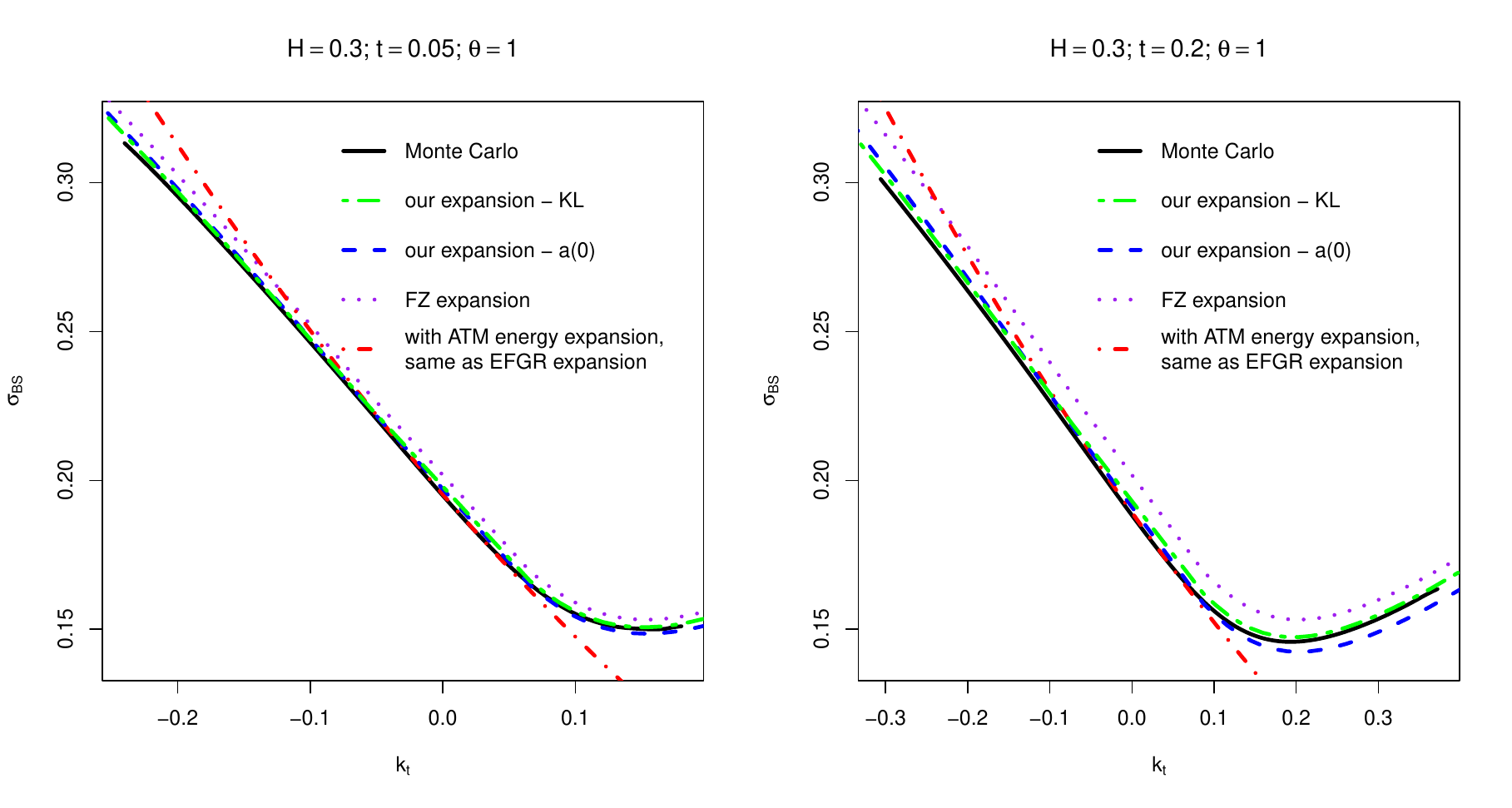}
	\caption{\label{fig:KL_time} 
Implied volatility smile approximations 
for the rBergomi model with parameters
$\theta=1,\sigma_0 = 0.2,
\eta = 1.5,
\rho = -0.7,
H = 0.3$, for expiry $t=0.05,\,0.2$. 
The Monte Carlo price is computed via the hybrid scheme for rBergomi in \cite{BLP15} with $\kappa = 2$, with $10^9$ simulations and $500$ time steps of length $t/500$.
The rate function is computed using the Ritz method in Section \ref{sec:ritz} with $N=8$ Haar basis functions,  the coefficient $a(x)$ is computed using the Karhunen-Loeve decomposition with  $N=300$ Haar basis functions (KL). We also compare with $a(x)$ expanded at $0$ ($a(x)\approx a(0)$).
}
    \end{center}
\end{figure}

Our goal in this section is to compare expansion \eqref{exp:implied_vol_KL} with other known implied volatility expansions under RoughVol.  
We consider:
\begin{itemize}
\item Implied volatility from Monte Carlo pricing, using the hybrid scheme for rBergomi in \cite{BLP15} with $\kappa = 2$ (note that a slight modification of the implementation is necessary for $\theta\neq 1$). 
\item  Our implied volatility expansion, where the term structure coefficient $a(x)$ is computed using KL, so that we have \eqref{exp:implied_vol_KL}, or where $a(x)$ is expanded at $0$, so that we have \eqref{exp:implied_variance_rate_function}.
\item The FZ expansion \eqref{bbf}. In \cite{forde2017asymptotics}, Forde and Zhang show that this asymptotics holds for volatilities of type $\sigma(t,\omega)=\sigma(\hat{W}_t)$, with no direct dependence on $t$, so this applies to \eqref{eq:funct:general:rbergomi} for $\theta=0$. However, as we have shown in \cite[Section 7.3]{friz2018precise1}, the same large deviation behavior holds when $\theta\neq 0$. Therefore, the FZ expansion gives the same asymptotic smile, independently of the choice of $\theta$.
\item Expansion \eqref{exp:implied_variance_rate_function}, with ATM expansion of $\Sigma$ as in \eqref{exp:implied_variance} (so, rate function is expanded as well). In case $\theta=1$,  one can check that this approximation is consistent with the expansion in \cite[Section 5]{euch2018short}, that we refer to as ``EFGR expansion''. These two mathematical results ar different, since log-strikes are in our case (large deviation regime) $k_t=xt^{1/2-H}$ and in \cite{euch2018short} (central limit regime) $k_t=xt^{1/2}$. However, when plotting for finite $k$ and $t$ the approximate implied volatility, the two curves are the same.
\end{itemize}
We first use the numerical methods detailed in next Section \ref{sec:projections} to compute $\Sigma(x)$ and $a(x)$. In Figure \ref{fig:KL_time} we display implied volatility smiles in the rBergomi model with $\theta=1$, for varying $t$, where the rate function is computed using the Ritz method in Section \ref{sec:ritz} and the coefficient $a(x)$ is computed using the KL decomposition from Section \ref{sec:KL}. For comparison, we also use approximation $a(x)\approx a_0$, and show  \eqref{exp:implied_variance_rate_function}. We notice that both implementations perform well, and the use of KL decomposition gives a better approximation of the right wing. On several simulations, this improvement of KL over expansion $a(x)\approx a_0$ is more evident when taking $\theta=1$, less when $\theta=0$.
Practically, implementation of the KL formula requires to approximate the infinite product \eqref{eq:CarlemanFredholm}, and we observed that for smaller values of $H$ the convergence of this product was much slower, requiring a prohibitively large number of basis functions, which is why we present these results for $H=0.3$.
We leave the numerically efficient implementation of the KL decomposition method for small values of $H$ as a topic for future research. In what follows we will consider the approximation $a(x)\approx a(0)$, which is faster while still producing accurate smiles.

\begin{figure}[ht!]
    \begin{center}
	\includegraphics[width=165mm]
	{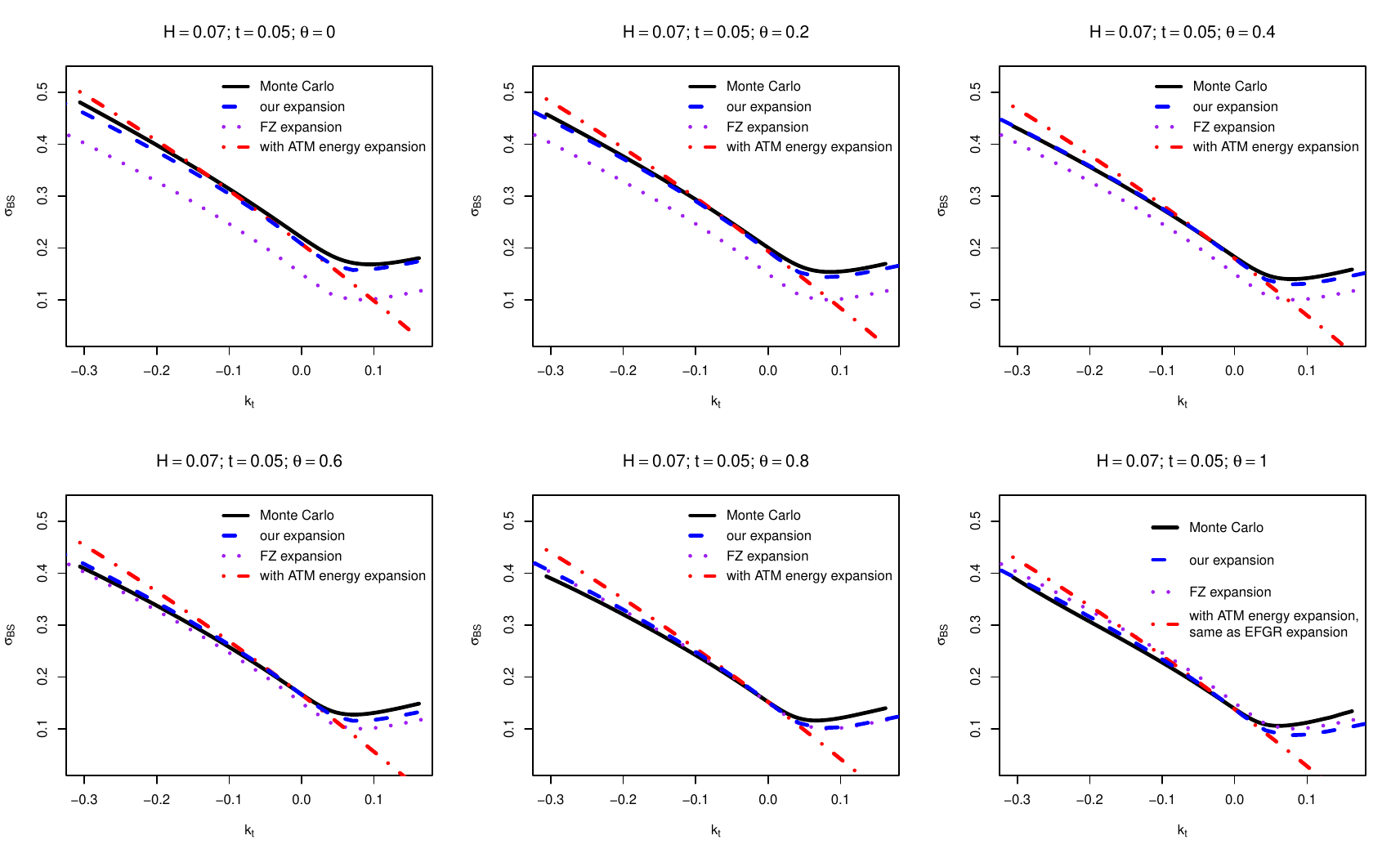}
	\caption{\label{fig:var:theta} 
Implied volatility smile approximation 
for the rBergomi model with parameters
$\sigma_0 = 0.15,
\eta = 1.8,
\rho = -0.78,
H = 0.07$, for expiry $t=0.05$. 
The Monte Carlo price is computed via the hybrid scheme for rBergomi in \cite{BLP15} with $\kappa = 2$, with $10^9$ simulations and $500$ time steps.
The rate function is computed using the Ritz method with $N=9$ Fourier basis functions.
}
    \end{center}
\end{figure}

First, in Figure \ref{fig:var:theta} we show implied volatilities under model \eqref{eq:vol:general:rbergomi}, with realistic parameters (close to the calibrated parameter to the SPX volatility on February 4, 2010, see \cite{bayer2016pricing}),  varying $\theta$ from $0$ to $1$. 
We note how our approximation is general enough to be applicable for any $\theta$, improving previous asymptotics in all cases. We also note a slight deterioration of the quality of the approximation in the right wing as $\theta \to 1$, that could be improved using KL to compute $a(x)$.

\begin{figure}[ht!]
    \begin{center}
	\includegraphics[width=165mm]
	{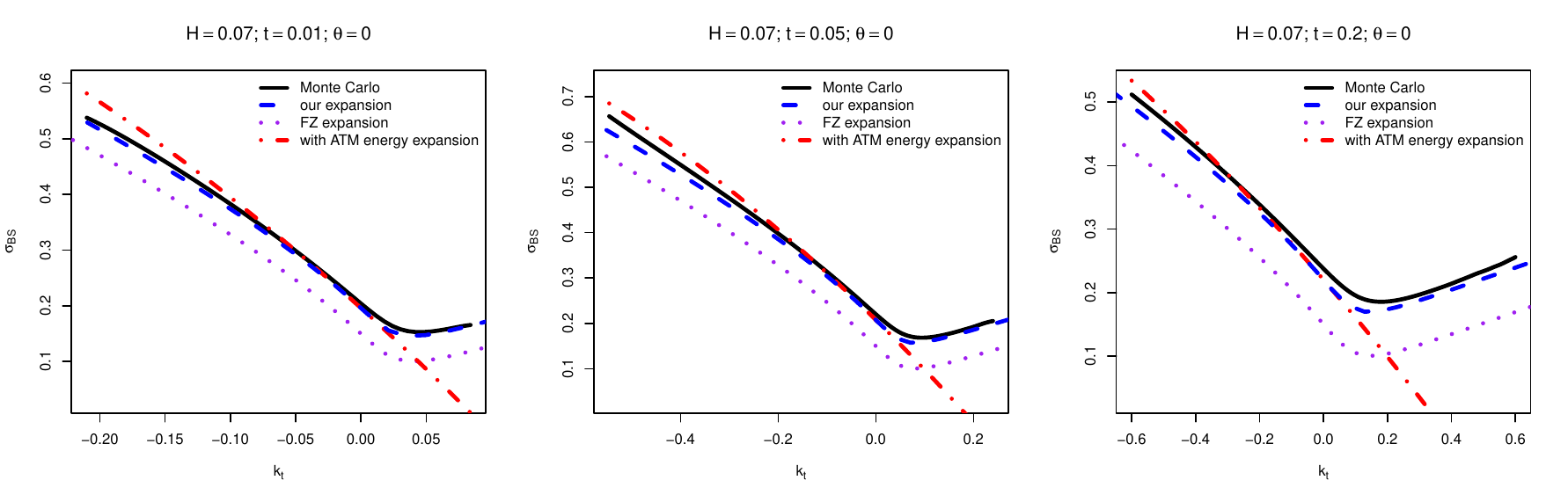}
	\caption{\label{fig:var:time} 
Implied volatility smile approximation 
for the rBergomi model with parameters
$\theta=0, \sigma_0 = 0.15,
\eta = 1.8,
\rho = -0.78,
H = 0.07$, for expiry $t=0.01,0.05,0.2$. 
The Monte Carlo price is computed via the hybrid scheme for rBergomi in \cite{BLP15} with $\kappa = 2$, with $10^9$ simulations and $500$ time steps of length $t/500$. The rate function is computed using the Ritz method with $N=9$ Fourier basis functions.
}
    \end{center}
\end{figure}

Then, instead of varying $\theta$, we fix $\theta=0$ and show in Figure \ref{fig:var:time} the comparison with the same approximations as before, when the expiry $t$ increases. We see how our expansion lifts the FZ expansion, improving the approximation of the Monte Carlo price. The difference between the two approximations is due to the term structure correction $a_0 t^{2H}$. Clearly, the effect of this correction becomes more evident as $t$ increases. On a number of numerical experiments, it is also clear that this correction becomes more and more important as $H\to 0$, not surprisingly since $t^{2H}$ is larger, for small $t$, when $H$ vanishes.

\begin{figure}[ht!]
    \begin{center}
	\includegraphics[width=165mm]
	{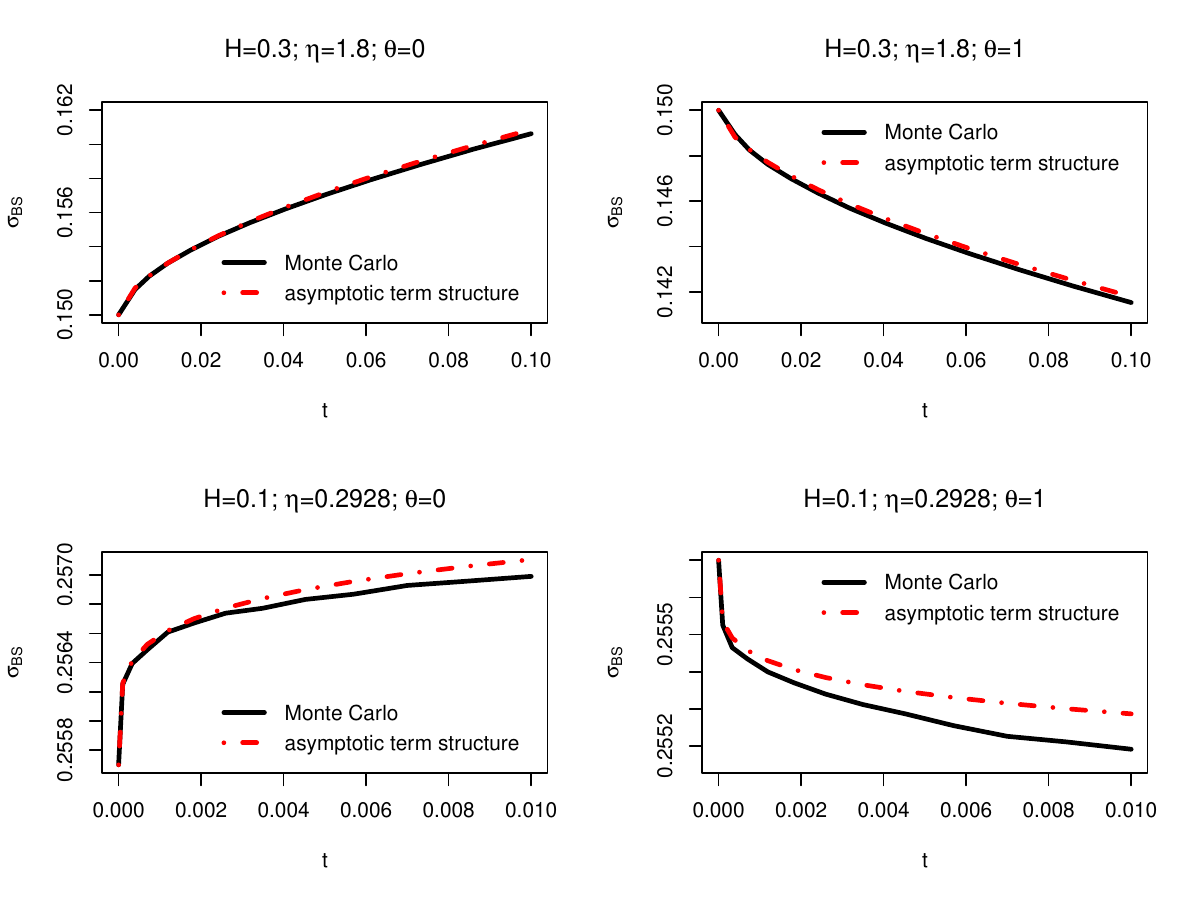}
	\caption{\label{fig:termstructure:time} 
	Term structure of volatility for the rBergomi model with parameters $\sigma_0 = 0.15,  \eta = 1.8, \rho = -0.78, H = 0.3$ (above) and with parameters $\sigma_0 = 0.2557,  \eta = 0.2928, \rho = -0.7571, H = 0.1$ (below). We plot ATM implied volatility as expiration time increases. We consider
	shorter expiries in the case of rougher trajectories (smaller Hurst parameter $H$; however, in this case we also take a smaller vol-of-vol parameter $\eta$).
	The Monte Carlo prices are computed via the hybrid scheme in \cite{BLP15} with $\kappa = 2$, with $10^9$ simulations and $500$ time steps. 
}
    \end{center}
\end{figure}

Now we check how our approximations behave as time increases. To do so, in Figure \ref{fig:termstructure:time} we show the ATM term structure of implied volatility, comparing ATM implied volatilities computed using Monte Carlo simulations and expansion \eqref{expansion:ATM:ivol}, for rBergomi with $\theta=0$ and $\theta=1$. 
We do so for parameters as in Figures \ref{fig:var:theta}  and \ref{fig:var:time}, with $H=0.3$, and for a different choice of parameters with $H=0.1$ and a smaller volatility of volatility $\eta$,  as in  \cite[Section 4]{bayer2017short}. The value of $\eta$ and $H$ affect the quality of the approximation, which is less accurate for $H$ very close to $0$ and $\eta>1$. On the other hand, as we show in Figure \ref{fig:termstructure:time}, for $H=0.3$ and $\eta>1$ or $H$ very close to $0$ and $\eta<1$ the short-time approximation is very good. This is consistent with the considerations on the interplay of $H$ and $\eta$ in \cite[Page 505]{euch2018short}.
We also see how the term structure is increasing in case $\theta=0$ and decreasing in case $\theta=1$. This is always the case: $a_0$ in \eqref{expansion-a} is always positive for $\theta=0$, always negative for $\theta=1$ (cf. Remark \ref{rem:empirical:term structure}). Also note that if the coefficient $\sigma_0$ were taken non-constant, the slope of the term structure would also be affected.

\begin{figure}[ht!]
    \begin{center}
	\includegraphics[width=105mm]
	{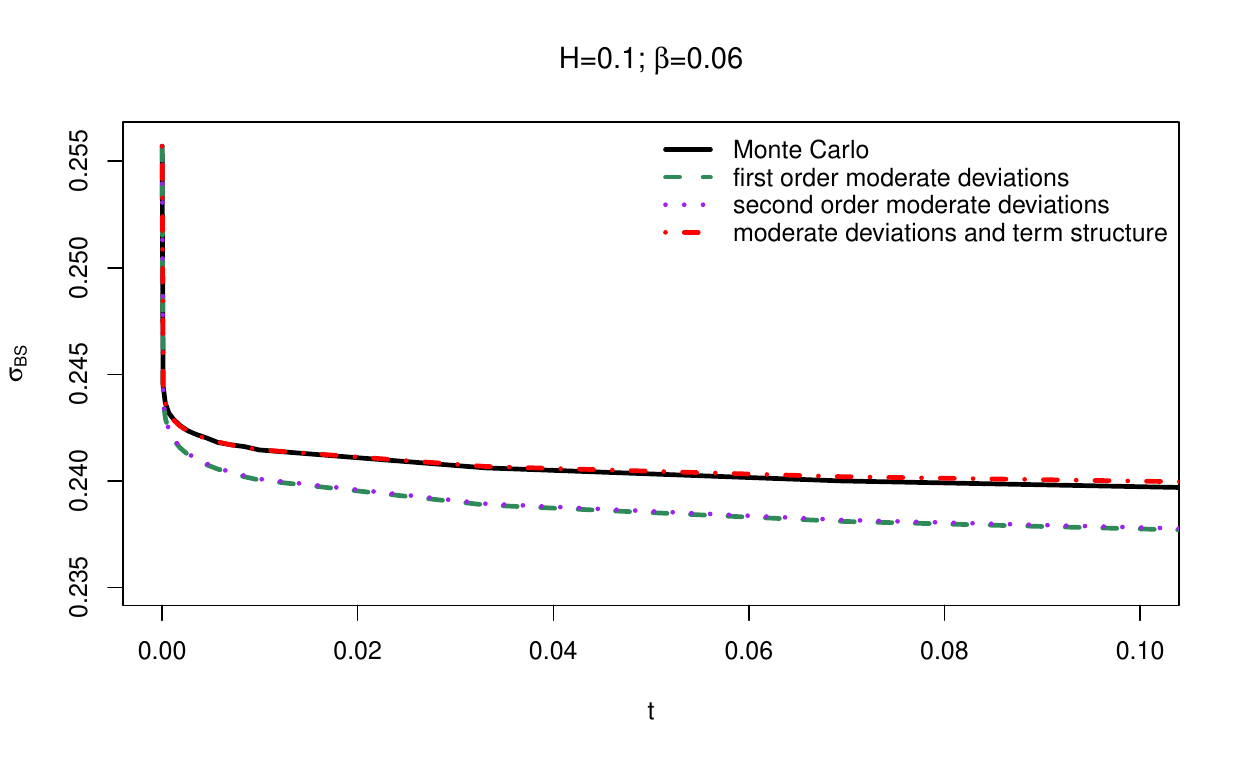}
	\caption{\label{fig:mod:dev} 
Moderate deviation with 	$\beta=0.06$ and $x=0.4$ (time varying log-strike $k_t= x t^{1/2-H+\beta}$)
of implied volatility in rBergomi model with 
$ \sigma_0 = 0.2557,
\eta = 0.2928, \rho = -0.7571, 
H = 0.1, \theta=0$. 
Simulation parameters:
                              $10^8$ simulation paths, $500$ time steps. Time interval $[0,0.1]$.
	}
   \end{center}
\end{figure}

Finally, as in Remark \ref{rem:moderate_deviations}, we consider moderate deviations. Figure \ref{fig:mod:dev}  is as in \cite[Figure 1]{bayer2017short}, the ``very rough'' case $H=0.1$ (which was the most problematic case in \cite{bayer2017short}). 
We are plotting, with $k_t=xt^{1/2-H+\beta}$, where $\beta=0.06$, the Monte Carlo implied volatility and its approximation
\[
\sigma(t,k_t)\approx \Sigma(0)+\Sigma'(0)x t^\beta + \frac{\Sigma''(0)}{2}x^2 t^{2\beta}+\frac{a_0}{2\sigma_0}t^{2H},
\]
considering terms up to the first order moderate deviation $t^\beta$, then up to the second order moderate deviation $t^{2\beta}$, and finally considering also the term structure $t^{2H}$.  We see how the term structure term improves the moderate deviation pricing. This also explains why, in \cite{bayer2017short}, the moderate deviation pricing gets worse as $H\downarrow 0$, since the distance of such price from the real (Monte Carlo) one is of order $t^{2H}$. We also see that using the second order moderate deviation actually does not improve much, and this follows from the fact that the curvature is almost $0$ with such choice of parameters (cf. Remark \ref{rem:moderate_deviations}). As for the term structure, the accuracy of the approximation formula based on moderate deviations gets worse as $\eta$ increases, for fixed $H$.

\begin{remark}
As mentioned above, Monte Carlo pricing is implemented using the hybrid scheme, which introduces a bias in the volatility process, while this process could be simulated exactly. However, in extensive simulations we find that the exact simulation scheme is more unstable for very short maturities, even with a $10^9$ trajectories and $500$ time steps. This is most likely due to the singularity of the kernel at $0$, which is what the hybrid scheme takes care of. On the other hand, with such large number of paths and fine discretisation, for larger maturities the two schemes display no visible difference.
Following these considerations, we used for our figures  Monte Carlo prices  simulated via the hybrid scheme in \cite{BLP15} .
\end{remark}

\begin{remark}
In \cite[Section 4.5]{forde2017asymptotics} asymptotics for model \eqref{eq:S:roughvolmodel} with volatility driven by a Mandelbrot-Van Ness fBm \eqref{fBM:VolterraRepresentation} are implemented. 
Without being completely rigorous, we have applied our expansion also in this case. We computed the $K$-functional numerically, as in this case no explicit formulas are available. Also in this case the term $a_0 t^{2H}$ lifts the smile, which gets closer to the real (Monte Carlo) implied volatility, for small $|x|$, with respect to the sole FZ expansion.
\end{remark}

\section{Computing the coefficients via projections}\label{sec:projections}

\subsection{Computing $\Sigma(x)$ using the Ritz method}\label{sec:ritz}

In order to use \eqref{exp:implied_vol_KL}, the first challenge is the computation of the rate function. A numerical approximation to $\Lambda$ can be obtained as described in \cite[Section 40]{gelfand_fomin}, using the Ritz method, as is done in \cite{forde2017asymptotics}. 
Natural choices for the orthornormal basis (ONB) $\{e_i\}_{i\geq 1}$ of $H^1$ are the Fourier basis,
\begin{equation}\label{eq:onb}
\dot{e}_1(s)= 1,\quad
\dot{e}_{2n}(s) =
\sqrt{2}\cos(2\pi n s ),\quad
\dot{e}_{2n+1}(s) =
\sqrt{2}\sin(2\pi n s ),\quad
\mbox{ for }n \in\N\setminus\{0\},
\end{equation}
or the Haar basis,
\begin{equation}\label{eq:onb:haar}
\dot{e}_1(s)= 1,\quad
\dot{e}_{2^k + l}(s) = 2^{k/2} \left(1_{\left[\frac{2l-2}{2^{k+1}},\frac{2l-1}{2^{k+1}}\right]}(s) - 1_{\left[\frac{2l-1}{2^{k+1}},\frac{2l}{2^{k+1}}\right]}(s) \right)
\mbox{ for }k  \geq 0, \; 1 \leq l \leq 2^k.
\end{equation}
We consider functions $h\in H^1$ with $h(0)=0$ so that
\[
\dot{h}(s)= \sum_{n=1}^N a_n \dot{e}_n(s),
\]
for $N\in \N$ fixed. Then we minimize
\[
\Lambda(x)
=
\frac{(x-\rho G(h))^2}{2\bar{\rho}^2 F(h)}
+\frac{\langle \dot{h},\dot{h} \rangle}{2},
\]
with
\begin{equation}\label{def:F:G}
\begin{split}
F(h) =\langle \sigma^2(\hat{h},0),1\rangle,\quad
G(h)=\langle \sigma(\hat{h},0),\dot{h}\rangle,
\end{split}
\end{equation}
over the Fourier coefficients $(a_n)_n$.
This representation of the energy function is also taken from \cite{forde2017asymptotics} (see notation in  \cite[Proposition 5.1]{bayer2017short}). The minimizing value for $\Lambda(x)$ is therefore our approximation for the energy and the corresponding function $\hat{h}^x$ is the approximate most likely path for the fBm $W^H$ associated with final condition $x$.

\subsection{A stochastic Taylor development}

The following stochastic Taylor expansion is sketched in \cite[Section 7.2]{friz2018precise1} for $\sigma(\omega,t)=\sigma(W^H_t)$. As discussed in Section \ref{sec:preliminary} and \cite[Section 7.3]{friz2018precise1}, our expansions can actually be carried out in the more general setting $\sigma(\omega,t)=\sigma(W^H_t,t^{2H})$. 
Under such volatility dynamics, the (rescaled)  log-price process is as in \eqref{eq:rescaled:log:price}.
As in \cite[Section 7.2]{friz2018precise1}, we can shift the dynamics via $\hat \eps (W, \bar W) \mapsto  (\hat \eps W + h , \hat \eps \bar W + \bar h) $, and apply Girsanov theorem in order to center Brownian fluctuations in the minimizer. Then, a stochastic Taylor expansion gives
\begin{eqnarray}\label{taylor:expansion}
\int_0^1\s \left( \hat\eps \hat W_t+\hat h_t,\hat \eps^2 t^{2H}\right)
d[ \hat\eps \tilde W + \tilde h ]_t
-\frac{\eps \, \hat\eps}{2} \int_0^1\sigma ^{2}\left( \hat\eps
\hat W_t+\hat h_t,\hat \eps^2 t^{2H}\right) dt \nonumber 
 \equiv  g_0 + \hat \eps g_1 (\omega) + \hat \eps^2 g_2 (\omega) + r_3(\omega) \ ,  \nonumber 
\end{eqnarray}%
where $r_3(\omega)$ is small\footnote{The precise control of this remainder is detailed in  \cite{friz2018precise1} and requires the sophisticated mathematical framework of regularity structures, that we do not intend to introduce in this paper. The interested reader  is referred to \cite{bayer2017regularity,friz2018precise1}.}, with
\begin{equation} \label{def:g1} 
g_1 
= \int_0^1 \sigma'(\hat{h}_s^x,0) \hat{W}_s d \tilde{h}_s^x 
+ \int_0^1 \sigma(\hat{h}_s^x,0) d \tilde{W}_s,
\end{equation}
(cf. \cite[Section 7.2]{friz2018precise1}) and 
\begin{equation}\label{def:g2}
g_2 =\begin{cases}
 \frac{1}{2}\int_0^1 \sigma''(\hat{h}_s^x,0) \hat{W}^2_s d\tilde{h}_s^x + \int_0^1\sigma'(\hat{h}_s^x,0) \hat{W}_s d \tilde{W}_s
 +
\int_0^1 \dot{\sigma}(\hat{h}_s^x,0) s^{2H} d\tilde{h}_s^x 
  & \mbox{ if } H<1/2, \\
 \frac{1}{2}\int_0^1 \sigma''(\hat{h}_s^x,0) \hat{W}^2_s d\tilde{h}_s^x + \int_0^1\sigma'(\hat{h}_s^x,0) \hat{W}_s d \tilde{W}_s
 +
\int_0^1 \dot{\sigma}(\hat{h}_s^x,0) s^{2H} d\tilde{h}_s^x 
 -\frac{1}{2} \int_0^1 \sigma^2(\hat{h}_s^x,0) ds
 & \mbox{ if } H= 1/2.
\end{cases}
\end{equation}
The following formula for $\Delta_2$ follows 
as \cite[Equation 7.5]{friz2018precise1}
\begin{equation}\label{def:Delta2}
\Delta_2 =\begin{cases}
 \frac{1}{2}\int_0^1 \sigma''(\hat{h}_s^x,0) \hat{V}^2_s d\tilde{h}_s^x + \int_0^1\sigma'(\hat{h}_s^x,0) \hat{V}_s d \tilde{V}_s, 
 +
\int_0^1 \dot{\sigma}(\hat{h}_s^x,0) s^{2H} d\tilde{h}_s^x 
  & \mbox{ if } H<1/2, \\
 \frac{1}{2}\int_0^1 \sigma''(\hat{h}_s^x,0) \hat{V}^2_s d\tilde{h}_s^x + \int_0^1\sigma'(\hat{h}_s^x,0) \hat{V}_s d \tilde{V}_s
 +
 \int_0^1 \dot{\sigma}(\hat{h}_s^x,0) s^{2H} d\tilde{h}_s^x 
 -\frac{1}{2} \int_0^1 \sigma^2(\hat{h}_s^x,0) ds
 & \mbox{ if } H= 1/2.
\end{cases}
\end{equation}
where we write 
\[
v_t = E[W_t g_1]/E[g_1^2],\quad 
\bar{v}_t =  E[\bar{W}_t g_1]/E[g_1^2],
\quad \tilde{v}_t=\rho v_t+\bar{\rho} \bar{v}_t,
\quad
 \hat{v}=K\dot{v}
\]
and
\[
\tilde{V}_t=\tilde{W}_t-\tilde{v}_t g_1 , \quad
\hat{V}_t=\hat{W}_t-g_1 \hat{v}_t.
\]

\subsection{Computing $a(x)$ using Karhunen-Loeve decomposition}\label{sec:KL}

Assume we are given $h^x$ computed by the Ritz method. Note then that $\bar{h}^x$ is obtained from $h^x$ via the following formula
\begin{equation} \label{eq:barh}
\dot{\bar{h}^x}(s) = \frac{x- \rho G(h^x)}{\bar{\rho} F(h^x)} \sigma(\hat{h}^x(s)),
\end{equation}
with $G(h), F(h) $ as in \eqref{def:F:G},  as can be seen by optimizing over $\bar{h}$ for fixed $h$ in the definition \eqref{ifzt} of the rate function.
Then we assume a Karhunen-Loeve (KL) decomposition of $(W,\bar{W})$:
\[ W = \sum_{i} \gamma_i e_i, \;\; \bar{W} = \sum_i \bar{\gamma_i} e_i,\]
where $\{e_i\}_i$ is the ONB in \eqref{eq:onb} or  \eqref{eq:onb:haar}
and $\gamma_i, \bar{\gamma}_i$ are i.i.d. standard Gaussians. 
This implies
\[ \hat{W} = \sum_{i} \gamma_i \hat{e}_i, \]
with $ \hat{e_i}(t)= (K * \dot{e}_i)_t$.
This yields 
\begin{equation*}
g_1 = \sum_{i} \cg_{i} \gamma_i +  \bar{\cg_{i}} \bar{\gamma_i}, 
\end{equation*}
where
\[
\begin{split}
 \cg_{i} &= \int_0^1 \sigma'(\hat{h}_s^x,0) \hat{e_i}(s) d \tilde{h}^x(s) + \rho \int_0^1 \sigma(\hat{h}_s^x,0) d e_i(s), \\
  \bar{\cg_{i}} &=  \bar{\rho} \int_0^1 \sigma(\hat{h}_s^x,0) d e_i(s).
\end{split}
\]
In particular 
\[ \sigma_x^2 = \sum_{i} \cg_{i}^2 +  \bar{\cg_{i}}^2.\]
Note then that 
\[ 
v(t) = \sum_i \cg_{i} e_i(t)  / \sigma_x^2,  \;\;\;\bar{v}(t) = \sum_i \bar{\cg_{i}} e_i(t)  / \sigma_x^2
,  \;\;\;
\hat{v}(t) = \sum_i \cg_{i} \hat{e}_i(t)  / \sigma_x^2,
\]
We then can write all the terms in $\Delta_2$ as follows.
We denote
\begin{equation*}
\alpha_{ij} = \int_0^1 \sigma''(\hat{h}_s^x,0) \hat{e_i}(s) \hat{e_j}(s) d\tilde{h}_s^x,
\quad
\beta_{ij} = \int_0^1\sigma'(\hat{h}_s^x,0) \hat{e_i} d e_j,
\end{equation*}
$\delta_{ij}={\bf 1}_{i=j}$ and  
 $\tilde{\cg}_i = \rho \cg_{i} + \bar{\rho} \bar{\cg}_i$.
Now, expanding \eqref{def:Delta2} with some long but standard computations we get to 
\[  \Delta_2 = \sum_{ij} (\gamma_i \gamma_j - \delta_{ij}) \eta_{0; ij} +  \gamma_i \bar{\gamma_j} \eta_{1; ij} +   (\bar{\gamma_i} \bar{\gamma_j} -\delta_{ij}) \eta_{2; ij}  + C =:  \Delta_2^{(2)}+C,\]
where
\begin{align*}
\eta_{0; ij}=& \frac 1 2 \alpha_{ij} - \frac{1}{\sigma_x^2}  \cg_{i} \sum_k \cg_k \alpha_{jk}  +  \frac{1}{2 \sigma_x^4}  \cg_{i} \cg_{j} \left( \sum_{k l } \cg_k \cg_l \alpha_{kl} \right)  + \rho \beta_{ij} - \frac{1}{\sigma_x^2}  \cg_{i} \sum_k \tilde{\cg}_k \beta_{jk}  \\
& - \frac{\rho}{\sigma_x^2}  \cg_{i} \sum_k \cg_k \beta_{kj} +  \frac{1}{\sigma_x^4} \cg_{i} \cg_{j} \left(\sum_{k,l} \cg_k \tilde{\cg}_l \beta_{kl} \right)
\\
\eta_{1; ij}=& \bar{\rho} \beta_{ij} - \frac{1}{ \sigma_x^2}  \bar{\cg}_j \sum_k \cg_k \alpha_{ik} + \frac{1}{2 \sigma_x^4} \cg_{i} \bar{\cg}_j \left(\sum_{k,l} \cg_k \cg_l \alpha_{k,l} \right)  - \frac{1}{ \sigma_x^2}  \bar{\cg}_j \sum_k \tilde{\cg}_k \beta_{ik} -  \frac{\bar{\rho}}{ \sigma_x^2} \cg_{i}  \sum_k \cg_k  \beta_{kj}
\\
& - \frac{{\rho}}{ \sigma_x^2} \bar{\cg}_j  \sum_k \cg_k  \beta_{ki}  + \frac{1}{\sigma_x^4} \cg_{i} \bar{\cg}_j  \left( \sum_{k} \sum_l \cg_k \tilde{\cg}_l \beta_{kl} \right)
\\
\eta_{2; ij}=&  \frac{1}{2\sigma_x^4}  \bar{\cg_{i}}  \bar{\cg_{j}} \left(\sum_{k,l} \cg_k \cg_l \alpha_{k,l}  \right) - \frac{ \bar{\rho}}{\sigma_x^2} \bar{\cg_{i}} \sum_k \cg_k \beta_{kj}  +  \frac{1}{\sigma_x^4}  \bar{\cg_{i}}  \bar{\cg_{j}} \left(\sum_{k,l} \cg_k \tilde{\cg}_l \beta_{k,l}  \right)
\end{align*}
and 
\begin{align*}
C=  & \int_0^1 \dot{\sigma}(\hat{h}_s^x,0) s^{2H} d\tilde{h}_s^x + \frac 1 2 \sum_i \alpha_{ii} - \frac{1}{2\sigma_x^2}  \sum_{i,k} \cg_{i}  \cg_k \alpha_{ik} -  \frac{1}{\sigma_x^2}\sum_{i,k} \cg_{i}  \tilde{\cg}_k \beta_{ik} .
\end{align*}
Recall that one has 
\[
A(x) =  e^{\Lambda'(x) C} \mathbb{E} \exp( \Lambda'(x) \Delta_2^{(2)}),
\]
where
\[\Lambda'(x) = \sgn(x)\sqrt{\frac{2 \Lambda(x)}{\sigma_x^2}},
\]
and since $\Delta_2^{(2)}$ is an element of the homogeneous Wiener chaos of order $2$, the expectation above can be computed as the Carleman-Fredholm determinant $\det_2(I-2 M)^{-1/2}$, where
 $M$ is the symmetric matrix 
\[ M =  \Lambda'(x) \left( \begin{array}{cc}\frac{1}{2} (\eta_0 + \eta_0^t)&\eta_1 \\ (\eta_1)^t & \frac{1}{2} (\eta_2 + \eta_2^t) \end{array} \right).\]
Namely one has
\begin{equation} \label{eq:CarlemanFredholm}
E \exp( \Lambda'(x) \Delta_2^{(2)}) = \Pi_{k \geq 0}  (1- 2\lambda_k)^{-1/2} e^{-\lambda_k} 
\end{equation}
where $(\lambda_k)_k$ are the eigenvalues of $M$ (note that the fact that all $\lambda_k < 1/2$ comes from the non-degeneracy assumption). This formula is a simple integral computation if $M$ is diagonal, and the general case follows by diagonalisation, cf e.g. \cite[Remark 5.5]{inahama2013} or \cite[p.78]{janson1997gaussian}.

Of course, in practice we consider approximations $W^N$, $\bar{W}^N$ obtained by truncating the sums to only keep indices $i\leq N$, where $N$ is fixed, so that all the sums above are then replaced by finite sums. One also needs to compute numerically the integrals appearing in the definition of the coefficients $\cg$, $\alpha$, $\beta$. We have found the Haar basis to be more convenient than the Fourier basis for this purpose since the $\hat{e}_i$'s have explicit expressions in that case. 
\section{Proofs}\label{sec:proofs}

\subsection{Energy expansion} 

\begin{lemma}[Fourth order energy expansion]
\label{expansion:J}
Consider a stochastic volatility model following dynamics \eqref{eq:S:roughvolmodel} and the associated energy function in \eqref{ifzt}.
Let $\Lambda(x)$ be the energy function in \eqref{ifzt}. Then
\begin{equation*}
\begin{split}
\Lambda(x)&= \frac{\Lambda''(0)}{2} x^2 +
\frac{\Lambda'''(0)}{3!} x^3+\frac{\Lambda^{(4)}(0)}{4!} x^4+O(x^5)\,\\
\end{split}
\end{equation*}
where
\begin{equation}\label{exp:J:2:3}
\Lambda''(0)=\frac{1}{\sigma_0^2}
,\quad \Lambda'''(0)=- 6 \frac{ \rho \sigma_0'}{\sigma_0^4} \langle K1,1 \rangle, 
\end{equation}
and
\begin{equation*}
\Lambda^{(4)}(0)
=12  \frac{(\sigma_0')^2}{\sigma_0^6}
\left\{
9 \rho^2 
 \langle K1,1\rangle^2
- \rho^2
 \langle (K1)^2 ,1\rangle  
-
 \langle (\bar{K}1)^2 ,1\rangle  
-
2 \rho^2 
\langle K1,\bar{K}1\rangle
\right\}
-12 \frac{\sigma_0''}{\sigma_0^5}   \rho^2
 \langle (K1)^2 ,1\rangle.  
\end{equation*}
\end{lemma}
\begin{remark}
In this lemma we expand the rate function $\Lambda(x)$, which has been studied first in \cite{forde2017asymptotics}. The second and third order terms in \eqref{exp:J:2:3} have been computed in \cite[Theorem 3.4]{bayer2017short}. In both these papers, the volatility function is supposed to be $\sigma(W^H_t)$, but adding the dependence $\sigma(W^H_t,t^{2H})$ does not change the large deviations behavior, meaning that the rate function is the same as the one of the model given by $\sigma(W^H_t,0)$. 
\end{remark}
\begin{proof}
We have the following development for the minimizer $h^x_\cdot$ in \eqref{ifzt}, for $x\rightarrow 0$:
\begin{equation}\label{dev:h}
h_t^x=\alpha_t x +\beta_t \frac{x^2}{2} +\gamma_t \frac{x^3}{6} +O(x^4),
\end{equation}
with
\[
\begin{split}
\alpha_t&=\frac{\rho}{\sigma_0} t , \quad\\
\beta_t&=2\frac{\sigma_0'}{\sigma_0^3}[\rho^2\langle K1,1_{[0,t]}\rangle+\langle K1_{[0,t]},1\rangle-3\rho^2 t \langle K1,1\rangle],
\end{split}
\]
where $\alpha,\beta$ have been also computed in \cite{bayer2017short}. We make here the ansatz that the expansion goes on one more order with $\gamma$, that we do not actually need to compute.
The existence of such $\gamma$ follows from the smoothness of $\sigma(\cdot,\cdot)$ (cf. \cite{friz2018precise1} and \cite[Section 5.2]{bayer2017short}).
We can compute, using $\langle K(K1),1\rangle = \langle K1,\bar{K}1\rangle$ and $\langle K(\bar{K}1),1\rangle =\langle (\bar{K}1)^2,1\rangle$,
\[
\begin{split}
K\dot{\beta}&=2\frac{\sigma_0'}{\sigma_0^3}[\rho^2 K(K1)+K(\bar{K}1) -3\rho^2 \langle K1,1\rangle K1] \ , \\
\langle K\dot{\beta},1\rangle& 
=2\frac{\sigma_0'}{\sigma_0^3}[\rho^2 \langle K1,\bar{K}1\rangle
+\langle (\bar{K}1)^2,1\rangle -3\rho^2 \langle K1,1\rangle^2] \ , \\
\langle K1,\dot{\beta}\rangle&=2\frac{\sigma_0'}{\sigma_0^3}[\rho^2 \langle (K1)^2,1\rangle+\langle K1,\bar{K}1 \rangle -3\rho^2 \langle K1,1\rangle^2] \ .
\end{split}
\]
We also have
\begin{equation}\label{exp:sigmahat}
\sigma(\hat{h}_s^x,0)=
\sigma_0+x \frac{\sigma_0'}{\sigma_0} \rho K1(s)+
\bigg( 
\frac{\sigma_0''}{\sigma_0^2} \rho^2 (K1)^2 (s) +\sigma_0' K\dot{\beta}(s)
\bigg)\frac{x^2}{2}+O(x^3) \ .
\end{equation}
We use now \eqref{def:F:G}  and compute
\begin{equation}
\begin{split}
F(h^x) &
=
\sigma_0^2+
x 2 \rho \sigma_0' \langle K1,1\rangle+
x^2 \bigg\{ \bigg(\bigg(\frac{\sigma_0'}{\sigma_0}\bigg)^2  + 
 \frac{\sigma_0''}{\sigma_0}  \bigg) \rho^2
 \langle (K1)^2 ,1\rangle +\sigma_0\sigma_0' \langle K\dot{\beta},1\rangle \bigg\}
+O(x^3) \ ,
\\
G(h^x)&
=
\rho x
+x^2  \bigg(
\frac{\sigma_0'}{\sigma_0^2} \rho^2 \langle K1,1\rangle+
\frac{\sigma_0}{2} \beta_1  \bigg)
+
x^3\bigg(
\frac{\sigma_0}{6} \gamma_1
+
\frac{\sigma_0''}{2\sigma_0^3} \rho^3 \langle(K1)^2,1\rangle  
\\&
\quad\quad
\quad\quad
+ \rho 
\frac{(\sigma_0')^2}{\sigma_0^4}\big[ 
(\rho^2+1)
\langle K1,\bar{K}1\rangle
+\rho^2 \langle (K1)^2,1 \rangle
+ \langle (\bar{K}1)^2,1 \rangle
-6\rho^2 \langle K1,1 \rangle^2\big]
\bigg)+O(x^3) \ ,
\end{split}
\end{equation}
from which we get
\[
\begin{split}
x-\rho G(h^x)&=
(1-\rho^2) x
-x^2  \rho
\frac{\sigma_0'}{\sigma_0^2} (1-\rho^2) \langle K1,1\rangle  
- 
x^3\rho
\bigg(
\frac{\sigma_0}{6} \gamma_1+
\frac{\sigma_0''}{2\sigma_0^3} \rho^3 \langle(K1)^2,1\rangle 
\\
&
+ \rho 
\frac{(\sigma_0')^2}{\sigma_0^4}\big[ 
(\rho^2+1)
\langle K1,\bar{K}1\rangle
+\rho^2 \langle (K1)^2,1 \rangle
+ \langle (\bar{K}1)^2,1 \rangle
-6\rho^2 \langle K1,1 \rangle^2\big]
\bigg)
+O(x^3) \ ,\\
\frac{1}{F(h^x)}
&=
\frac{1}{\sigma_0^2}-
x 2 \rho \frac{\sigma_0'}{\sigma_0^4}
 \langle K1,1\rangle-
x^2 \bigg\{ 
\frac{\sigma_0''}{\sigma_0^5}   \rho^2
 \langle (K1)^2 ,1\rangle  
 \\
&
+
\bigg(\frac{\sigma_0'}{\sigma_0^3}\bigg)^2 \bigg(
\rho^2
 \langle (K1)^2 ,1\rangle  
+2
 \langle (\bar{K}1)^2 ,1\rangle  
 +2\rho^2  \langle K1,\bar{K}1\rangle 
 -
10\rho^2 
  \langle K1,1\rangle^2\bigg)\bigg\} 
+O(x^3) \ ,
\\
\frac{(x-\rho G(h^x))^2}{1-\rho^2}&
=
(1-\rho^2) x^2
-x^3 2 
  \rho
\frac{\sigma_0'}{\sigma_0^2} (1-\rho^2) \langle K1,1\rangle  
+
x^4 
\bigg[
\rho^2
\frac{(\sigma_0')^2}{\sigma_0^4} (1+11\rho^2) \langle K1,1\rangle^2  
-
\frac{\sigma_0''}{\sigma_0^3} \rho^4 \langle(K1)^2,1\rangle 
\\
&
- \frac{\rho\sigma_0}{3} \gamma_1
- 2\rho^2 
\frac{(\sigma_0')^2}{\sigma_0^4} 
\big[
(\rho^2+1)
\langle K1,\bar{K}1\rangle
+\rho^2 \langle (K1)^2,1 \rangle
+ \langle (\bar{K}1)^2,1 \rangle
\big]
\bigg)\bigg] 
+O(x^5)
\ ,
\\
\frac{(x-\rho G(h^x))^2}{2(1-\rho^2) F(h^x)}
&= (\dots)x^2+(\dots)x^3 -
x^4\frac{\rho
}{6\sigma_0} \gamma_1
+x^4
\bigg\{
-\frac{\rho^2}{2}\frac{\sigma_0''}{\sigma_0^5}   
 \langle (K1)^2 ,1\rangle 
-2\rho^2 
\frac{(\sigma_0')^2}{\sigma_0^6}\langle K1,\bar{K}1\rangle
\\
&
- \frac{(\sigma_0')^2}{\sigma_0^6} \frac{\rho^4+\rho^2}{2}
 \langle (K1)^2 ,1\rangle  
- \frac{(\sigma_0')^2}{\sigma_0^6} 
 \langle (\bar{K}1)^2 ,1\rangle   
+
\frac{(\sigma_0')^2}{\sigma_0^6} \frac{3}{2} \rho^2  (5-\rho^2)
 \langle K1,1\rangle^2
 \bigg\}  +O(x^5)
 \ .
\end{split}
\]
We also have, from 
\eqref{dev:h}
\[
\langle \dot{h}^x,\dot{h}^x \rangle=
\dots +
x^4\bigg(\frac{\rho}{3\sigma_0}\gamma_1+ 
\big(\frac{\sigma_0'}{\sigma_0^3}\big)^2\big[ 
\rho^4\langle (K1)^2,1\rangle
+
\langle (\bar{K}1)^2,1\rangle
+2\rho^2\langle K1,\bar{K}1\rangle+3\rho^4\langle K1,1 \rangle^2
-6 \rho^2 \langle K1,1 \rangle \langle \bar{K}1,1 \rangle
\big]
\bigg)+O(x^5) \ .
\]
Now we write, from \cite[Proposition 5.1]{bayer2017short},
\[
\Lambda(x)
=
\frac{(x-\rho G(h^x))^2}{2\bar{\rho}^2 F(h^x)}
+\frac{\langle \dot{h}^x,\dot{h}^x \rangle}{2} \ ,
\]
and use the expansions above for the two summands. The fourth order expansion of $\Lambda(x)$ follows.
\end{proof}

\subsection{Proof of Lemma \ref{expansion:A}}

Let us take $x\neq 0$.
\\
STEP 1:
We first need to expand $\bar{h}^x_t$ in \eqref{ifzt}, for small $x$ (an expansion of $h^x$ was computed in \cite{bayer2017short}). We write
\begin{equation}\label{eq:itomap}
\Phi_1(W,\bar{W})=X_1
\end{equation}
for the It\^o map associated with the RoughVol model \eqref{eq:roughvolmodel}.
Computing the Frechet derivative of $\Phi_1$ with respect to the second component at $\h=(h,\bar{h})$ in the direction $f$ we get (cf. \eqref{def:g1})
\begin{equation}\label{frechet}
\langle D \Phi_1(\h) ,(0,f) \rangle
=
\langle D_2 \Phi_1(\h) ,f \rangle
=
\frac{d}{d\delta} \Phi_1(h,\bar{h}+\delta f) = \bar\rho \int_0^1 \sigma(\hat{h},0) df
\ .
\end{equation}	
From the first order optimality condition \cite[Appendix B]{friz2018precise1}, we get that for $\h^x$ minimizer and any $\f$ in the Cameron-Martin space $H^1$,
\[
\langle   \h^x_t, \f_t \rangle_{H^1}=\Lambda'(x) \langle D \Phi_1 ,\f \rangle.
\]
Let $f$ be the second component of $\f$. Using \eqref{frechet} we get
\[
\int_0^1 \dot{\bar{h}}^x_t \dot{f}_t dt =\langle   \bar{h}^x_t, f_t \rangle_{H^1}=\Lambda'(x) \langle D_2 \Phi_1 ,f \rangle =\bar{\rho} \Lambda'(x) \int_0^1 \sigma(\hat{h}_t^x,0)  \dot{f}_t dt.
\]
Now, from \eqref{exp:J:2:3} we derive that, for $x\rightarrow 0$, 
\begin{equation}\label{derLambda}
\Lambda'(x)=\frac{x}{\sigma_0^2} - 3 x^2 \frac{ \rho \sigma_0'}{\sigma_0^4} \langle K1,1 \rangle +O(x^3)
\ .
\end{equation}
We get
\[
\bar{h}^x_t = x \frac{\bar{\rho}}{\sigma_0}t +O(x^2) \ .
\]
We also have
\begin{equation}\label{expansion:h}
h^x_t = x \frac{{\rho}}{\sigma_0}t +O(x^2),\quad
\tilde{h}^x_t =\rho h^x_t+\bar{\rho}\bar{h}^x_t=  x \frac{t}{\sigma_0}+O(x^2),\quad
\hat{h}^x_t =
(K\dot{h}^x)_t =
x \frac{\rho}{\sigma_0} K1(t) \ ,
\end{equation}
and
\[
\sigma(\hat{h}^x,0)=
\sigma_0 +x \frac{\rho \sigma_0'}{\sigma_0} K1+O(x^2), \quad
\sigma'(\hat{h}^x,0)=
\sigma_0' +x \frac{\rho \sigma_0''}{\sigma_0} K1+O(x^2), \quad
\dot{\sigma}(\hat{h}^x,0)=
\dot{\sigma}_0 +O(x) \ .
\]
STEP 2: We recall here, from \cite{friz2018precise1}, the definition of some quantities needed to compute $A(x)$. Let $g_1$ be as in \eqref{def:g1} and let us write $\sigma_x^2=Var(g_1)$ for its variance. We recall, again from  \cite[Equation (6.3)]{friz2018precise1}, $\sigma_x^2= 2\Lambda(x)/\Lambda'(x)^2$, 
from which we get
\begin{equation}\label{expansionsigmax}
\sigma_x^2=\sigma_0^2+4\rho\sigma_0'\langle K1,1\rangle x +O(x^2) \ .
\end{equation}
From \eqref{def:g1} we define and compute
\begin{equation}\label{eq:vs}
\begin{split}
v_t& = \frac{ E[W_t g_1]}{E[g_1^2]} =
\frac{1}{\sigma_x^2}\bigg(
\rho \int_0^t  \sigma(\hat{h}_s^x,0) ds + \int_0^1 \sigma'(\hat{h}_s^x,0) K1_{[0,t]}(s) d \tilde{h}_s^x
 \bigg),\\
\bar{v}_t &= \frac{ E[\bar{W}_t g_1]}{E[g_1^2]} = \frac{1}{\sigma_x^2}\bar{\rho} \int_0^t \sigma(\hat{h}_s^x,0) ds.
\end{split}
\end{equation}
(Note that $v, \bar{v}$ are in the Cameron-Martin space). 
From \eqref{Aabstract} we have that $A(x)$ in Theorem \ref{thm:main:smalltime} is  
\[
A(x)=E\left[\exp(\Lambda' (x) \Delta_2)\right]
\ ,
\] 
where $\Delta_2$ is given in \eqref{def:Delta2}.
\\
STEP 3:
We can expand now such quantity, for $x\rightarrow 0$ and we get
\begin{equation}\label{exp:taylor:A}
\begin{split}
A(x)
=1 + x\Lambda''(0)E\left[\Delta_2^0\right] +x^2 \bigg(\frac{\Lambda'''(0)E\left[\Delta_2^0\right]
+\Lambda''(0)^2  E\left[(\Delta_2^0)^2\right] 
}{2}   + 
 \Lambda''(0)  E\left[\partial_x \big|_{x=0}\Delta_2\right]
\bigg)
+
O(x^3) \ ,
\end{split}
\end{equation}
where $\Delta_2^0$ denotes $\Delta_2 \big|_{x=0}$. The statement of the theorem follows from the computation of the quantities in \eqref{exp:taylor:A}.
\\
STEP 4: 
We compute
\[
\dot{v}_t=
\frac{1}{\sigma_x^2}\bigg(
\rho \sigma(\hat{h}_t^x,0) + \int_t^1 \sigma'(\hat{h}_s^x,0) K(s,t) d \tilde{h}_s^x
 \bigg)
 \ ,
 \]
and we obtain, also using \eqref{eq:vs},
\begin{equation}\label{eq:v}
\begin{split}
\sigma_x^2 v_t
=
\rho \sigma_0 t +
x  \frac{\sigma_0'}{\sigma_0} \big(\rho^2 
\langle K1,1_{[0,t]}\rangle 
+ \langle K1_{[0,t]},1\rangle 
\big)+ O(x^2)
\ ,\\
\sigma_x^2 \tilde{v}_t
=
\sigma_0 t +
x  \frac{\sigma_0'}{\sigma_0} \rho 
\big(\langle K1,1_{[0,t]}\rangle+\langle K1_{[0,t]},1\rangle 
\big) + O(x^2)
\ ,
\\
\sigma_x^2 \hat{v}_t
=  \rho \sigma_0 
K1(t)
+
x  \frac{\sigma_0'}{\sigma_0} \big(\rho^2 K(K1)(t)
+ K(\bar{K}1)(t)
\big)
+O(x^2)
\ ,
\end{split}
\end{equation}
where we have used
$
\langle K1_{[0,t]},1\rangle =\int_0^t \bar{K}1(u)du.
$
We have
\[
\sigma_x^4 \hat{v}_t d\tilde{v}_t
=
\rho \sigma_0^2 
K1(t) dt
+
x  \sigma_0'  \Big(
\rho^2 K(K1)(t)
+ K(\bar{K}1)(t)
+ \rho^2 K1(t) \big(
K1(t)+\bar{K}1(t) 
\big)  
\Big)
dt \ .
\]
Putting together the previous expressions and using $\langle K(K1),1\rangle=\langle K1,\bar{K}1\rangle$ and $\langle K(\bar{K}1),1\rangle=\langle (\bar{K}1)^2,1 \rangle$ we get
\begin{equation}\label{eq:vdv}
\begin{split}
\sigma_x^4 \int_0^1 \hat{v}_t d\tilde{v}_t
&=
\rho \sigma_0^2 
\langle K1,1 \rangle
+
x  \sigma_0'  \bigg( \rho^2 \langle(K1)^2,1\rangle   + \langle (\bar{K}1)^2,1\rangle 
+2 \rho^2 \langle K1, \bar{K}1\rangle  
\bigg)
+O(x^2)
\ ,
\\
\sigma_x^4 \int_0^1 K1(t)\hat{v}_t d\tilde{v}_t
&=
\rho \sigma_0^2 
\langle (K1)^2,1\rangle
+O(x) 
\ ,
\\
\sigma_x^4 \int_0^1 \hat{v}_t^2 dt &=\rho^2 \sigma_0^2 \langle (K1)^2,1 \rangle +O(x) \ .
\end{split}
\end{equation}
This implies, toghether with \eqref{expansionsigmax},
\begin{equation}\label{dxsigmaint}
\partial_x\bigg(\sigma_x^2 \int_0^1 \hat{v}_t d\tilde{v}_t\bigg)
=
\frac{\sigma_0'}{\sigma_0^2}
 \bigg( 2\rho^2\langle K1,\bar{K}1 \rangle
+
 \rho^2 \langle(K1)^2,1\rangle   
 + \langle (\bar{K}1)^2,1\rangle 
-4 \rho^2
\langle K1,1 \rangle^2 
\bigg)
\ .
\end{equation}
We can now compute
\[
\begin{split}
E\int_0^1  \bigg(\hat{V}_s d \tilde{V}_s\bigg)
=
E\int_0^1  \hat{W}_s d \tilde{W}_s +E[g_1^2] \int_0^1  \hat{v}_s d \tilde{v}_s
-
E \bigg[g_1 \bigg( \int_0^1 \hat{W}_s d \tilde{v}_s + \hat{v}_s d \tilde{W}_s \bigg)\bigg]
\ ,
\end{split}
\]
where
\[
\begin{split}
E\big[ g_1\int_0^1 \hat{W}_s d \tilde{v}_s \big]
=
\int_0^1\int_0^s K(s,u) dE\big[ g_1 W_u \big] d \tilde{v}_s
=
\sigma_x^2
\int_0^1\hat{v}_s d \tilde{v}_s \ ,\\
E\big[ g_1 \int_0^1 \hat{v}_s d \tilde{W}_s \big]
=
\int_0^1 \hat{v}_s d E\big[g_1 \tilde{W}_s \big]
= \sigma_x^2
 \int_0^1 \hat{v}_s d \tilde{v}_s \ ,
\end{split}
\]
so that
\[
\begin{split}
E\int_0^1  \bigg(\hat{V}_s d \tilde{V}_s\bigg)
= -\sigma_x^2\int_0^1  \hat{v}_s d \tilde{v}_s
\ .
\end{split}
\]
We also compute
\[\begin{split}
& \int_0^1  E[\hat{V}_s^2] ds =
\langle K^2 1, 1\rangle -
\sigma_x^2\int_0^1 \hat{v}_s^2 ds \ , \\
& \int_0^1 
 K1(s) E[\hat{V}_s d \tilde{V}_s] 
=
- \sigma_x^2 \int_0^1 
 K1(s) \hat{v}_s d \tilde{v}_s \ ,
 \end{split}
 \]
and all these quantities can be expansionded in $x$ using \eqref{eq:vdv}.
Now we use \eqref{def:Delta2} to write, in the case $H<1/2$
\begin{equation}\label{ED02}
E \Delta_2^0
 =
 \sigma'_0
E \int_0^1 \hat{V}^0_s d \tilde{V}^0_s
=- \rho\sigma'_0
\langle K1,1 \rangle.
\end{equation}
Moreover, using \eqref{expansion:h},
\[
\partial_x \big|_{x=0} \int_0^1 \dot{\sigma}(\hat{h}^x_s,0) s^{2H} d \tilde{h}^x_s
=
\frac{\dot\sigma_0}{(2H+1)\sigma_0}
\ .
\]
Now, also using \eqref{def:Delta2} and \eqref{eq:vdv} we get
\[\begin{split}
\partial_x  E \Delta_2 \big|_{x=0}
&=
  \frac{\sigma''_0}{2\sigma_0}\int_0^1  E [ (\hat{V}^0_s)^2] ds
+ \rho \frac{\sigma''_0}{\sigma_0}\int_0^1   
 K1(s)  E[ \hat{V}_s^0 d \tilde{V}_s^0 ]
+\sigma'_0\int_0^1   \partial_{x} E[\hat{V}_s d \tilde{V}_s] \big|_{x=0}
+\frac{\dot\sigma_0}{(2H+1)\sigma_0}
\\
&=
\frac{\sigma''_0}{\sigma_0}
\bigg(\frac{\langle K^2 1, 1\rangle}{2}
-
\frac{3}{2}\rho^2  \langle (K1)^2,1 \rangle
\bigg)
\\
&
-
\bigg(\frac{\sigma'_0}{\sigma_0}\bigg)^2
 \bigg( 2\rho^2\langle K1,\bar{K}1 \rangle
+
 \rho^2 \langle(K1)^2,1\rangle   
 + \langle (\bar{K}1)^2,1\rangle 
-4 \rho^2
\langle K1,1 \rangle^2 
\bigg)
+\frac{\dot\sigma_0}{(2H+1)\sigma_0}
\ .
\end{split}\]
STEP 5:
We need now to compute
\[
E[(\Delta_2^0)^2]=
(\sigma_0')^2
E \bigg(\int_0^1 \hat{V}^0_s d \tilde{V}^0_s\bigg)^2,
\]
where (using definitions and \eqref{eq:v})
\[
\hat{V}^0_s = \hat{W}_s - \rho K1(s) \tilde{W}_1 \quad \mbox{
and } \quad
d \tilde{V}^0_s = d \tilde{W}_s -  \tilde{W}_1 ds \ .
\]
We can rewrite
\[
\begin{split}
\int_0^1 \hat{V}^0_s ds \tilde{W}_1
= 
\bar{\rho} \int_0^1 \bar{K}1(u) \tilde{W}_u d B _u
+
\int_0^1
\bigg(
\rho\big(\bar{K}1(u)-2\langle K1, 1\rangle \big)
\tilde{W}_u
+ \int_0^u \bar{K}1(s)  dW_s 
\bigg)
d\tilde{W}_u
.
\end{split}
\]
and, differentiating the product $\int_0^1 K1(u) d\tilde{W}_u, \tilde{W}_1$
\[
\begin{split}
&\int_0^1 \hat{V}_s^0 d \tilde{V}^0_s
= 
- \bar{\rho} \int_0^1 \bar{K}1(u) \tilde{W}_u d B _u
+
\int_0^1
\bigg(\hat{V}_u^0- \int_0^u \bar{K}1(s)  dW_s +
\rho\big(2\langle K1, 1\rangle- \bar{K}1(u) \big)
\tilde{W}_u
\bigg)
d\tilde{W}_u \\
&= 
- \bar{\rho} \int_0^1 \bar{K}1(u) \tilde{W}_u d B _u
 -\rho \int_0^1 K1(u) du\\
&+
\int_0^1
\bigg(\hat{W}_u
- \int_0^u \bar{K}1(s)  dW_s +
\rho\big(2\langle K1, 1\rangle- \bar{K}1(u) -K1(u)\big)
\tilde{W}_u
-\rho\int_0^u K1(s)d\tilde{W}_s
\bigg)
d\tilde{W}_u
\end{split}
\]
with $\tilde{W}$ independent of $B$. Therefore, by It\^o isometry,
\[
\begin{split}
&E\bigg[\bigg(
\int_0^1 \hat{V}^0_s d \tilde{V}^0_s\bigg)^2\bigg]
=
\bar{\rho}^2 \int_0^1 \bar{K}1(u)^2 u d u 
+\rho^2 \bigg(\int_0^1 K1(u)du\bigg)^2
\\
&+
E\int_0^1 
\bigg(\hat{W}_u
- \int_0^u \bar{K}1(s)  dW_s +
\rho\big(2\langle K1, 1\rangle- \bar{K}1(u) -K1(u)\big)
\tilde{W}_u
-\rho\int_0^u K1(s)d\tilde{W}_s
\bigg)
^2
du.
\end{split}
\]
We can apply again It\^o isometry to compute the last expectations, and
\[
\begin{split}
E\bigg[\bigg(
\int_0^1 \hat{V}^0_s d \tilde{V}^0_s\bigg)^2\bigg]
&=
\rho^2\int_0^1 \int_0^u 
 \big(K(u,s)-\bar{K}1(s) +
2\langle K1, 1\rangle- \bar{K}1(u) - K1(u)-K1(s)
\big)^2 d s du \\ 
&
+\rho^2 \bigg(\int_0^1 K1(u)du\bigg)^2
+
\bar{\rho}^2 \int_0^1 \int_0^u 
(K(u,s)-\bar{K}1(s))^2  ds du 
 + 
\bar{\rho}^2 \int_0^1 \bar{K}1(u)^2 u d u.
  \end{split}
\]
At this point it is a (long) calculus excercise (noting $\langle \bar{K}1,1 \rangle = \langle K1,1 \rangle$) to show that 
\begin{equation}\label{EVdV}
\begin{split}
E\bigg[\bigg(
\int_0^1 \hat{V}^0_s d \tilde{V}^0_s\bigg)^2\bigg]
=
\rho^2(
3
\langle K1, 1\rangle^2
- \langle (K1)^2,1 \rangle
- 2 \langle K1,\bar{K}1 \rangle
 )
+
\langle K^2 1, 1\rangle-\langle (\bar{K}1)^2,1 \rangle
  \end{split}
\end{equation}
\\
STEP 6: 
Substituting in \eqref{exp:taylor:A} we get
\begin{equation*}
\begin{split}
A(x)
&=1 - x \frac{
\rho \sigma_0'}{\sigma_0^2} \langle K1,1 \rangle
+x^2 \bigg\{
\frac{ (\sigma_0')^2}{\sigma_0^4} 
\bigg( 3\rho^2 \langle K1,1 \rangle^2
+ \frac{1}{2} E \bigg[ \Big(
\int_0^1 \hat{V}^0_s d \tilde{V}^0_s\Big)^2 \bigg]
\bigg) + 
 \frac{\sigma''_0}{\sigma_0^3}
\bigg(\frac{\langle K^2 1, 1\rangle}{2}
-
\frac{3}{2}\rho^2  \langle (K1)^2,1 \rangle
\bigg)\\
 &
-
\frac{(\sigma'_0)^2}{\sigma_0^4}
 \bigg( 2\rho^2\langle K1,\bar{K}1 \rangle
+\rho^2
\langle (K1)^2,1\rangle
+
\langle (\bar{K}1)^2,1\rangle
-4 \rho^2
\langle K1,1 \rangle^2 
\bigg)+
\frac{\dot{\sigma}_0}{(2H+1)\sigma_0^3} \bigg\}
+
O(x^3)\\
=
&1 - x \frac{
\rho \sigma_0'}{\sigma_0^2} \langle K1,1 \rangle
+x^2 \bigg\{
\frac{ (\sigma_0')^2}{\sigma_0^4} 
\bigg(   
\frac{ \langle K^2 1, 1\rangle }{2}- \frac{3}{2}\langle (\bar{K}1)^2,1 \rangle
+
\rho^2
\Big(
\frac{17}{2}
\langle K1, 1\rangle^2
- \frac{3}{2}\langle (K1)^2,1 \rangle
- 3 \langle K1,\bar{K}1 \rangle \Big)
 \bigg)
\\
 & + 
 \frac{\sigma''_0}{\sigma_0^3}
\bigg(\frac{\langle K^2 1, 1\rangle}{2}
-
\frac{3}{2}\rho^2  \langle (K1)^2,1 \rangle
\bigg)
+
\frac{\dot{\sigma}_0}{(2H+1)\sigma_0^3} \bigg\}
+
O(x^3)
\end{split}
\end{equation*}
and we get Theorem \ref{expansion:A:rough}.
\\
STEP 7: When $H=1/2$, $\Delta_2$ in \eqref{def:Delta2} has an additional summand. Let us write 
\[
\tilde{\Delta}_2 =  \frac{1}{2}\int_0^1 \sigma''(\hat{h}_s^x,0) \hat{V}^2_s d\tilde{h}_s^x + \int_0^1\sigma'(\hat{h}_s^x,0) \hat{V}_s d \tilde{V}_s
 +
\int_0^1 \dot{\sigma}(\hat{h}_s^x,0) d\tilde{h}_s^x,
\]
so that $\tilde{\Delta}_2$ has the same expression as $\Delta_2$ in the rough case $H<1/2$. For $H=1/2$ we can write
 \begin{equation}
\Delta_2 = \tilde{\Delta}_2 -\frac{1}{2} \int_0^1 \sigma^2(\hat{h}_s^x,0) ds,
\end{equation}
so that
\begin{equation}
\Delta_2^0 = \tilde{\Delta}^0_2-\frac{\sigma_0^2}{2} 
\end{equation}
and, using \eqref{exp:sigmahat}
\begin{equation}
\partial_x\big|_{x=0} {\Delta}_2 = \partial_x \big|_{x=0} \tilde{\Delta}_2 - 
\sigma'_0 \rho \langle K1,1\rangle
\end{equation}
Now, $A(x)$ in Theorem \ref{thm:main:smalltime} is  
\[
A(x)=e^x E\left[\exp(\Lambda' (x) {\Delta}_2)\right]
\] 
with $\Delta_2$ as above. Expanding in $x$ we find
\begin{equation}
\begin{split}
A(x)
&=1 + x(\Lambda''(0)E[{\Delta}_2^0]+1) +x^2 \bigg(\frac{\Lambda'''(0)E[{\Delta}_2^0]
+
E\left[(\Lambda''(0){\Delta}_2^0+1)^2\right] 
}{2}   + 
 \Lambda''(0)  E\left[\partial_x \big|_{x=0}{\Delta}_2\right]
\bigg)
+
O(x^3) \\
&=1 + x \Lambda''(0)E[ \tilde{\Delta}_2^0]
+x^2 \bigg(\frac{\Lambda'''(0)E[{\tilde{\Delta}}_2^0]
+\Lambda''(0)^2  E[({\tilde{\Delta}}_2^0)^2] 
}{2}   + 
 \Lambda''(0)  E[\partial_x \big|_{x=0}{\tilde{\Delta}}_2 ]  \bigg)
 + \frac{x}{2} + \frac{x^2}{8}+
O(x^3)
\end{split}
\end{equation}
(we have used \eqref{exp:J:2:3} and \eqref{ED02}).

\subsection{Proof of Theorem \ref{expansion:atm:ivol}}
A Taylor expansion gives
\[
\Sigma(x)=\frac{x}{\sqrt{2 \Lambda(x)}}= 
\frac{1}{\sqrt{\Lambda''(0)}}\bigg(1
-
\frac{\Lambda'''(0)}{6 \Lambda''(0)} x
+
\frac{\Lambda'''(0)^2-\Lambda''(0)\Lambda^{(4)}(0)}{24 \Lambda''(0)^2}
x^2
\bigg)+O(x^3).
\]
The explicit expressions for the three terms now follow from Lemma \ref{expansion:J}. 
Let us compute $a(x)$.
The rate function is quadratic and $\Lambda(0)=\Lambda'(0)=0$. Then, using Taylor developments of $\Lambda$ and $x\rightarrow \frac{1}{1+x}$ we get
\[
\frac{2\Lambda(x)}{x\Lambda'(x)}=1- \frac{x}{6} \frac{\Lambda'''(0)}{ \Lambda''(0)} + \frac{x^2}{12}\left\{ \bigg(\frac{\Lambda'''(0)}{\Lambda''(0)}\bigg)^2 
- \frac{\Lambda^{(4)}(0)}{\Lambda''(0)} \right\}+ O(x^3)
=1
+\sqrt{\Lambda''(0)} \bigg(x \Sigma'(0)
+ x^2 \Sigma''(0) \bigg)
+ O(x^3)
\]
From Lemma \ref{expansion:J},
\[
\frac{2\Lambda(x)}{x\Lambda'(x)}=1+ x \frac{\rho \sigma_0' \langle K1,1\rangle}{ \sigma_0^2} +
x^2
\sqrt{\Lambda''(0)}\Sigma''(0)
 + O(x^3)
\]
and, with $A(x)$ given in Lemma \ref{expansion:A:rough}, we have when $H<1/2$
\[
\begin{split}
\frac{2A(x) \Lambda(x)}{x\Lambda'(x)}=1+ x^2 
\bar{a}_0
 + O(x^3)
\end{split}
\]
with
\[
\begin{split}
\bar{a}_0&=\frac{(\sigma_0')^2}{\sigma_0^4} C_{K,\rho}+
 \frac{\sigma''_0}{\sigma_0^3}
\bar{C}_{K,\rho}
+
\frac{\dot{\sigma}_0}{(2H+1)\sigma_0^3}
+
\sqrt{\Lambda''(0)} v'' (0)
-
\frac{\rho^2 (\sigma_0')^2 \langle K1,1\rangle^2}{ \sigma_0^4}\\
&=\frac{(\sigma_0')^2}{\sigma_0^4} \frac{D_{K,\rho}}{2}+
 \frac{\sigma''_0}{\sigma_0^3}
\frac{ \bar{D}_{K,\rho}}{2}+
\frac{\dot{\sigma}_0}{(2H+1)\sigma_0^3},
\end{split}
\]
with $ D_{K,\rho}
, \bar{D}_{K,\rho},
$ defined in \eqref{def:D}.
Now, as a consequence of Lemma \ref{expansion:J}, we have
\[
\frac{x^2}{2\Lambda(x)^2}\sim \frac{2}{\Lambda''(0)^2 x^2}=\frac{2\sigma_0^4}{ x^2}
\]
and the expansion of $a(x)$ follows.
When $H=1/2$,
\[
\begin{split}
\frac{2A(x) \Lambda(x)}{x\Lambda'(x)\exp(x/2)}=1+ x^2 
\bar{a}_0
 + O(x^3)
\end{split}
\]
with
\[
\begin{split}
\bar{a}_0
&=\frac{(\sigma_0')^2}{\sigma_0^4} \frac{ D_{K,\rho} }{2}+
 \frac{\sigma''_0}{\sigma_0^3}
\frac{\bar{D}_{K,\rho}}{2} +
\frac{\dot{\sigma}_0}{(2H+1)\sigma_0^3}+
\frac{\rho \sigma_0'}{2\sigma_0^2}\langle K1,1\rangle.
\end{split}
\]
We conclude as in the case $H<1/2$.

\subsection{Proof of Theorem \ref{th:moderate_deviations}}
The call asymptotics is a corollary of 
Theorem \ref{thm:main:smalltime},
 taking into consideration that
\[
\Lambda(x_t)=\sum_{i=2}^{n} \frac{\Lambda^{(i)}(0)}{i!} x^i t^{i\beta} +O(t^{(n+1)\beta})
\]
and that $O(t^{(n+1)\beta-2H})\rightarrow 0$ under $ \beta \in(\frac{2H}{n+1}, \frac{2H}{n}]$.
Recall $\Lambda''(0)=\sigma_0^{-2}$ and the first statement follows.

Let us write $\alpha=2H-2\beta$,
$\delta=1/2+2H-2\beta$,
$\gamma=1/2-H+\beta$ and $
\M(t,x)=
\sum_{i=3}^{n} \frac{\Lambda^{(i)}(0)}{i!} x^i t^{i\beta-2H}
$.
We intend to apply \cite[Corollary 7.1, Equation (7.2)]{gaolee}, where $G_-(k,u)$ denotes $\sqrt{2}(\sqrt{u+k}-\sqrt{u})$
and $V$ denotes $\sqrt{t}\sigma_{BS}$. To do so, we notice that
\begin{equation}\label{eqq1}
G^2_-(k,u)=
\frac{k^2}{2u}+o\big(\frac{k^2}{u^2}\big)
\end{equation}
when $k\downarrow 0$ and $u\rightarrow \infty$.  In the notation of  \cite{gaolee}, we have
\[
L_t=-\log c(t,k_t),
\]
and $G$ will be computed for $u=L_t-\frac{3}{2}\log L_t+\log(\frac{k_t}{4\sqrt{\pi}})$, 
so  let us compute
\[
\begin{split}
&L_t-\frac{3}{2}\log L_t+\log(\frac{k_t}{4\sqrt{\pi}})\\
&=
\frac{{x^2}}{2\sigma_0^{2}t^{\alpha}}+\M(t,x) - \log \frac{\sigma_0^3}{{x^2}\sqrt{2\pi}}
- \delta \log t-\frac{3}{2}\log L_t+\log\big(\frac{k_t}{4\sqrt{\pi}}\big)+o(1)
\end{split}
\]
and take care of the logarithmic terms in $t$. For $t\downarrow 0$, 
\[
\begin{split}
- \delta \log t-\frac{3}{2}\log L_t+\log\big(\frac{k_t}{4\sqrt{\pi}}\big)
&= (- \delta+\frac{3}{2}\alpha + \gamma) \log t  -\frac{3}{2}\log (\frac{x^2}{2\sigma_0^{2}})
+\log\big(\frac{x}{4\sqrt{\pi}}\big)+o(1)
\\
&= -\frac{3}{2}\log (\frac{x^2}{2\sigma_0^{2}})
+\log\big(\frac{x}{4\sqrt{\pi}}\big)+o(1)
\end{split}
\]
So
\begin{equation}\label{eqq2}
L_t-\frac{3}{2}\log L_t+\log\big(\frac{x}{4\sqrt{\pi}}\big)= 
\frac{x^2}{2\sigma_0^{2} t^{\alpha}} + \M(t,x)  +o(1)
\end{equation}
Equations \eqref{eqq1} and \eqref{eqq2} tell us that
\begin{equation}\label{eqq3}
\begin{split}
&\frac{1}{t}G^2_-\big(k_t,L_t-\frac{3}{2}\log L_t+\log\big(\frac{k_t}{4\sqrt{\pi}}\big)\big)
=
\sigma_0^2
\frac{1}{
 1+\frac{2\sigma_0^2 \M(t,x)}{x^2}t^{\alpha}
 +o(t^{\alpha}) 
 } +o(t^{\alpha})
\end{split}
\end{equation}
The proof now boils down to writing the development of this factor using the Taylor developement of $\frac{1}{1+u}$, with $u=2\sigma_0^2 \M(t,x)t^{\alpha}/x^2
 +o(t^{\alpha})$.
We have, for  $j\in \N$,
\[
u^j
=
\left(
\frac{2\sigma_0^2\M(t,x)}{x^2}t^{\alpha}
\right)^j +o(t^{\alpha})
\]
using $\M(t,x)^{p-1} t^{j\alpha}=
o(t^{\alpha})$ for $j\geq p\geq 1$.
Also notice $u^{n-1}=O((\M(t,x)t^\alpha)^{n-1})=o(t^\alpha)$ because $ \beta \in(\frac{2H}{n+1}, \frac{2H}{n}]$. We have
\begin{equation}\label{eqq4}
\begin{split}
\frac{1}{1+u}&=\sum_{j=0}^{n-2}(-1)^ju^j+O(u^{n-1})=\sum_{j=0}^{n-2}(-1)^j \left(
\frac{2\sigma_0^2 \M(t,x)}{{x^2}}t^{\alpha}
\right)^j 
+ o(t^{\alpha})
\end{split}
\end{equation}
So from \eqref{eqq3} and \eqref{eqq4}
\begin{equation}\label{eqq5}
\begin{split}
\frac{1}{t}G^2_-(k_t,L_t-\frac{3}{2}\log L_t +\log(\frac{k_t}{4\sqrt{\pi}}))
=
\sum_{j=0}^{n-2}
(-1)^j 2^j
 \sigma_0^{2(j+1)}
 \left(
\frac{\M(t,x)}{{x^2}}t^{\alpha}
\right)^j
+ o(t^{\alpha})
\end{split}
\end{equation}

We apply now \cite[Corollary 7.1, Equation (7.2)]{gaolee}:
\[
\left|
\frac{1}{t}G^2_-(k_t,L_t-\frac{3}{2}\log L_t +\log(\frac{k_t}{4\sqrt{\pi}}))
-\sigma_{BS}^2(k_t)
\right| = o\big(\frac{k_t^2}{tL_t^2}\big)=o\big(t^{\alpha}\big)
\]
and obtain expansion \eqref{exp:imp:vol}.

\appendix

\section{Fractional Brownian motion}\label{sec:fBM}
The fBM is a ``rough'' continuous-time Gaussian process in that, depending on a parameter $H\in(0,1)$, its trajectories are locally H\"older continuous of any order strictly less than $H$. Unlike classical BM, the increments of fBm are not independent if $H\neq 1/2$.
The fBM was introduced for the first time by Mandelbrot and Van Ness in \cite{Mandelbrot1968} as the following stochastic integral, for $t \ge 0$:
\[
Z^H_t = c_H\bigg[\int_{-\infty}^t (t-s)^{H-1/2}dZ_s \,- \,\int_{-\infty}^0 (-s)^{H-1/2 }dZ_s\bigg],
\]
where $Z$ is a BM and  $c_H=\big(\int_0^\infty[(1+s)^{1/2-H}-s^{1/2-H}]^2 ds+\frac{1}{2H}\big)^{1/2}$.
Such process is Gaussian with covariance 
\begin{equation}\label{cov:fbm}
E[Z^H_t Z^H_s]=\frac{1}{2}(|t|^{2H}+|s|^{2H}-|t-s|^{2H}).
\end{equation} 
It can also be represented as a Volterra integral on the interval $[0,t]$:
\begin{equation}\label{fBM:VolterraRepresentation}
Z^H_t = \int_{0}^t  K_H(s,t) dB_s,
\end{equation}
with $K_H$ as in \cite{Nua06} or \cite[Section 3.1]{forde2017asymptotics}).
One can consider the following variant of fBM, known as \textit{Riemann-Liouville process} \cite{Mandelbrot1968},  introduced in 1953 by L\'{e}vy. This process is also represented as Volterra integral as
 \begin{equation}\label{eq:RLfbm}
 \hat{B}^H_t = \int_{0}^t K(t,s) dB_s,
 \end{equation}
with a simpler kernel
\begin{equation}\label{kernel:RLfBM}
 K(t,s) = \sqrt{2H} (t-s)^{H-1/2} , \mbox{ for }\quad H \in (0,1).
\end{equation}
 It is still self-similar, but stationarity of increments does not hold.
Moreover, the covariance structure is more complicated than \eqref{cov:fbm}. It can be expressed using hypergeometric functions (see \cite[Lemma 4.1]{bayer2017short}). The $K$-functionals that we find in our expansion can be computed in this case as
\begin{equation}\label{explicit:fractional}
\begin{split}
\langle K1,1\rangle
&=\frac{\sqrt{2H}}{(H+1/2)(H+3/2)} \\
\langle K^2 1, 1\rangle
&=\frac{1}{2H+1}\\
\langle (K1)^2,1 \rangle
=
\langle (\bar{K}1)^2,1 \rangle
&
=
\frac{H}{(H+1)(H+1/2)^2}
\\
\langle K1,\bar{K}1 \rangle
&=
\frac{2H}{(H+1/2)^2}\beta(H+3/2,H+3/2)
\end{split}
\end{equation}
where $\beta$ is the beta function.
In the case $K\equiv 1$ the fBM driving the volatility is actually a BM and we are back to the classical setting of a diffusive Markovian volatility. In this case our expansions can be compared e.g. to  \cite{MS03,MS07}.

\bibliographystyle{abbrv}
\bibliography{roughvol}
\end{document}